\documentclass[journal]{IEEEtran}
\usepackage{amsmath,amsfonts, amssymb}
\usepackage{bbm}
\usepackage{algorithmic}
\usepackage{algorithm}
\usepackage{array}
\usepackage{subcaption}
\usepackage{textcomp}
\usepackage{stfloats}
\usepackage{url}
\usepackage{xcolor}
\usepackage{physics}
\usepackage{verbatim}
\usepackage{graphicx}
\usepackage{cite}

\newcommand{\rev}[1]{\color{black}{{#1}}}
\newcommand{\revv}[1]{\color{black}{{#1}}}

\newcommand{\bea}{\begin{eqnarray}}
\newcommand{\eea}{\end{eqnarray}}

\newcommand{\inp}[2]{\left\langle {#1} \middle| {#2} \right\rangle}
\newcommand{\kb}[1]{\ket{{#1}}\!\bra{{#1}}}

\newcommand{\calY}{\mathcal{Y}}

\newcommand{\calH}{\mathcal{H}}

\newtheorem{theorem}{Theorem}
\newtheorem{corollary}[theorem]{Corollary}
\newtheorem{lemma}[theorem]{Lemma}
\newtheorem{proposition}[theorem]{Proposition}
\newtheorem{remark}{Remark}

\newtheorem{definition}{Definition}

\begin{document}

\title{Quantum Advantage in Locally Differentially \\ Private
Hypothesis Testing} 

\author{Seung-Hyun Nam,~\IEEEmembership{Member,~IEEE}, Hyun-Young Park,~\IEEEmembership{Member,~IEEE},
\\ Si-Hyeon Lee,~\IEEEmembership{Senior Member,~IEEE}, and Joonwoo Bae,~\IEEEmembership{Member,~IEEE}

\thanks{© 2026 IEEE. Personal use of this material is permitted. Permission from IEEE must be obtained for all other uses, in any current or future media, including reprinting/republishing this material for advertising or promotional purposes, creating new collective works, for resale or redistribution to servers or lists, or reuse of any copyrighted component of this work in other works.}
\thanks{This paper was presented in part at 2025 IEEE International Symposium on Information Theory (ISIT) \cite{namQuantumAdvantageLocallyISIT2025} and  at 13th Beyond IID in Information Theory.}
\thanks{S.-H. Nam is with the Information \& Electronics Research Institute, Korea Advanced Institute of Science and Technology (KAIST), Daejeon, South Korea (e-mail: shnam@kaist.ac.kr).}
\thanks{H.-Y. Park, S.-H. Lee, and J. Bae are with the School of Electrical Engineering, KAIST, Daejeon 34141, South Korea  (e-mail: phy811@kaist.ac.kr; sihyeon@kaist.ac.kr; joonwoo.bae@kaist.ac.kr). (Corresponding Author: Si-Hyeon Lee and Joonwoo Bae)}
}

\maketitle

\begin{abstract}
    We consider a private hypothesis testing scenario, including both symmetric and asymmetric testing, based on classical data samples.
    The utility is measured by the error exponents, namely the Chernoff information and the relative entropy, while privacy is measured in terms of classical or quantum local differential privacy.
    In this scenario, we show a quantum advantage with respect to the optimal privacy-utility trade-off (PUT) in certain cases.
    Specifically, we focus on distributions referred to as smoothed point mass distributions, along with the uniform distribution, as hypotheses.
    We then derive upper bounds on the optimal PUTs achievable by classical privacy mechanisms, which are tight in specific instances.
    To show the quantum advantage, we propose a particular quantum privacy mechanism that achieves better PUTs than these upper bounds in both symmetric and asymmetric testing, {\revv specifically under stringent privacy constraints and small discrete data alphabet sizes ranging from $3$ to $9$.}
    The proposed mechanism consists of a classical-quantum channel that prepares symmetric informationally complete (SIC) states, followed by a depolarizing channel.
\end{abstract}


\section{Introduction}\label{sec:intro}
\IEEEPARstart{I}{n} statistical inference, there is a risk of private information being leaked through the data to be collected \cite{narayananRobustDeanonymizationLarge2008a, fredriksonModelInversionAttacks2015, dworkExposedSurveyAttacks2017, geipingInvertingGradientsHow2020, rigakiSurveyPrivacyAttacks2023}.
The leakage of private information can be bounded by perturbing the data with a privacy mechanism that satisfies a privacy constraint.
One of the most representative privacy constraints is the local differential privacy (LDP) \cite{kasiviswanathanWhatCanWe2011a,duchiLocalPrivacyStatistical2013a, dworkAlgorithmicFoundationsDifferential2013}, which modifies the differential privacy (DP) \cite{dworkDifferentialPrivacy2006} to operate in the local model.
However, as implied by the data-processing inequality (DPI), applying a privacy mechanism degrades the accuracy of inference, referred to as the utility.
Consequently, one of the fundamental problems in private statistical inference is to characterize the optimal trade-off between the leakage of private information and the utility, known as the optimal \emph{privacy-utility trade-off} (PUT).

In {\rev the} classical setting, the optimal PUT has been exactly or approximately characterized for various inference tasks \cite{yeOptimalSchemesDiscrete2018, parkExactlyOptimalCommunicationEfficient2024, duchiLocalPrivacyStatistical2013a, asiOptimalAlgorithmsMean2022a, sheffetLocallyPrivateHypothesis2018, asoodehContractionLocallyDifferentially2024, canonneStructureOptimalPrivate2019, kairouzExtremalMechanismsLocal2016b}.
Among them, we focus on the locally differentially private hypothesis testing.
In this task, data providers produce their data after applying privacy mechanism satisfying the LDP constraint, and a data collector aims to infer the true hypothesis that the distribution of the raw data follows.
Hypothesis testing can be classified into two types: symmetric testing, where the utility is defined based on the average error probability, and asymmetric testing, where the utility corresponds to the trade‑off between type I and type II error probabilities.
Previous works \cite{sheffetLocallyPrivateHypothesis2018, asoodehContractionLocallyDifferentially2024, canonneStructureOptimalPrivate2019, kairouzExtremalMechanismsLocal2016b} studied the optimal PUT in symmetric or asymmetric testing, and especially, \cite{kairouzExtremalMechanismsLocal2016b} showed that an extremal mechanism achieves the optimal PUT in asymmetric testing exactly.

The notions of DP and LDP {\rev have} been extended to quantum systems, referred to as quantum DP (QDP) and quantum LDP (QLDP), respectively, and have been studied over the past decade \cite{zhouDifferentialPrivacyQuantum2017a, aaronsonGentleMeasurementQuantum2019, yoshidaClassicalMechanismOptimal2020, duQuantumDifferentiallyPrivate2022, watkinsQuantumMachineLearning2023, angrisaniUnifyingFrameworkDifferentially2023, guanDetectingViolationsDifferential2023, farokhiQuantumPrivacyHypothesisTesting2023, hircheQuantumDifferentialPrivacy2023, nuradhaQuantumPufferfishPrivacy2024, guanOptimalMechanismsQuantum2024, chengSampleComplexityLocally2024, yoshidaMathematicalComparisonClassical2025, nuradhaContractionPrivateQuantum2025, chengInvitationSampleComplexity2025, angrisaniQuantumDifferentialPrivacy2025}.
In a general quantum system, each data provider holds quantum data as their raw data and perturbs it into a quantum state.
Within this general scenario, \cite{chengSampleComplexityLocally2024, nuradhaContractionPrivateQuantum2025, chengInvitationSampleComplexity2025, angrisaniQuantumDifferentialPrivacy2025} considered locally differentially private hypothesis testing and studied the optimal PUT.
More specifically, \cite{angrisaniQuantumDifferentialPrivacy2025} analyzed the error exponent in asymmetric testing, and \cite{chengSampleComplexityLocally2024, nuradhaContractionPrivateQuantum2025, chengInvitationSampleComplexity2025} analyzed the sample complexity in both symmetric and asymmetric testing.

While the general quantum scenario, in which each data provider holds quantum data, is certainly meaningful, we instead focus on the case where they hold classical data, which is closer to currently prevalent practical situations. 
In this case, a fundamental question is whether there is a \emph{quantum advantage} in private hypothesis testing with respect to the optimal PUT, that is, whether replacing an optimal classical privacy mechanism with a quantum privacy mechanism can improve the PUT.
Regarding this question, the previous work \cite{yoshidaClassicalMechanismOptimal2020} provided a partial negative answer that there is no quantum advantage whenever the original classical data is binary.
Later, \cite{yoshidaMathematicalComparisonClassical2025} showed that quantum advantage exists in the optimal PUT for certain utility metrics related to the right logarithmic derivative (RLD) Fisher information of specific one-parametric families of quantum states.
Although the results in \cite{yoshidaMathematicalComparisonClassical2025} do not directly address a specific operational scenario for such utility metrics, they motivate the exploration of quantum advantage in private statistical inference.
To this end, we show for the first time that quantum advantage exists in private statistical inference in specific cases.
Our results open up the possibility of achieving a better PUT in private statistical inference systems through the use of a quantum privacy mechanism.

Specifically, we consider both symmetric and asymmetric hypothesis testing scenarios under the LDP or QLDP constraint, where the original data is classical and the utility is measured by error exponents.
{\revv We show that a quantum advantage exists in the optimal PUTs for a specific class of hypotheses, explicitly establishing this advantage under stringent privacy constraints and for small discrete data alphabet sizes ranging from $3$ to $9$.
To demonstrate this advantage, we propose a particular quantum privacy mechanism and derive upper bounds on the optimal PUTs achievable by classical privacy mechanisms.
The proposed mechanism can exploit both noise and the non-orthogonality of quantum states to manipulate indistinguishability.
While the use of noise is common to both LDP and QLDP, the intrinsic quantum indistinguishability of non-orthogonal states cannot be demonstrated in classical models \cite{PhysRevX.8.011015, PRXQuantum.3.030337}.}

{\revv In more detail,} we consider specific hypotheses involving smoothed point masses and the uniform distribution.
These hypotheses not only facilitate the analytical treatment of the PUTs, which is generally challenging, but also capture practical operational scenarios such as user preference surveys, as discussed in Section~\ref{sec:prob_formulation}.

The proposed quantum privacy mechanism first prepares a pure quantum state corresponding to its classical input, and then {\revv applies} depolarizing noise.
Here, the pure states to be prepared form a subset of symmetric informationally complete (SIC) states, and the amount of depolarizing noise is optimized to achieve the best PUT as possible. 

To derive upper bounds on the optimal PUTs achievable by classical LDP mechanisms, we show that every LDP mechanism can be simulated by a composition of an extremal mechanism \cite{kairouzExtremalMechanismsLocal2016b} followed by a post-processing.
Combining with the DPI, we get the upper bounds on the optimal PUTs by maximizing the utilities over extremal mechanisms.
Moreover, we show that these bounds are tight in certain cases by proving that block design mechanisms \cite{parkExactlyOptimalCommunicationEfficient2024} achieve the upper bounds.
We note that our results in characterizing the optimal PUTs achievable by classical LDP mechanisms also make their own contribution to the literature on classical private statistical inference.

The rest of this paper is organized as follows.
We first introduce preliminaries on QLDP and hypothesis testing in Section~\ref{sec:preliminaries}.
In Section~\ref{sec:prob_formulation}, we define locally differential private hypothesis testing scenario, PUT, and quantum advantage.
Section~\ref{sec:main} summarizes our main results on the quantum advantage.
To establish these results, we propose a certain QLDP mechanism in Section~\ref{sec:Q_mech} and, in Section~\ref{sec:CPUT}, characterize upper bounds on classical PUTs and show that these bounds are tight in some cases.
We prove the quantum advantage in Section~\ref{sec:Q_adv_pf} and provide the remaining proofs in Section~\ref{sec:pfs}.
Section~\ref{sec:conc} concludes the paper with a discussion on future work.

\section{Preliminaries}\label{sec:preliminaries}

In general, we consider both symmetric and asymmetric hypothesis testing scenarios, where classical data is perturbed into a quantum state to mitigate the leakage of private information.
To address these scenarios, we provide preliminaries on QLDP and both symmetric and asymmetric hypothesis testing.
Classical scenarios can be regarded as a specific instance of the general consideration.

\subsection{Notations}
Throughout, let $\calH_d$ denote a $d$-dimensional Hilbert space.
The set of quantum states on a Hilbert space $\calH$ is denoted by $\mathcal{D}(\calH)$.
A measurement is represented by a positive operator-valued measure (POVM) $\{\Lambda_y\}_{y \in \calY}$ where $\Lambda_y \geq 0$ and $\sum_{y \in \calY} \Lambda_y = I$, where $I$ is the identity operator.
For a vector $q$ of length $d$, we write $\mathrm{diag}(q)$ for the $d \times d$ diagonal matrix whose diagonal entries are $q_x$ for $x=1,\cdots, d$.
For $v \in \mathbb{N}$, let $[v] := \{1,\ldots,v\}$ and $[0:v]:=\{0\}\cup[v]$.
We write $\mathbbm{1}$ for a vector or matrix with all entries equal to $1$ whose subscript denotes its dimension, and $\delta_{h,x}$ denote the Kronecker delta.
For two real valued functions $f(\epsilon)$ and $g(\epsilon)$, we define $f(\epsilon) \overset{\epsilon\rightarrow 0}\approx g(\epsilon)$ to mean that $\lim_{\epsilon \rightarrow 0} f(\epsilon)/g(\epsilon) = 1$.

\subsection{Quantum local differential privacy}
A data provider holding its own data $X \in \mathcal{X}$ can mitigate privacy leakage by perturbing it into a quantum state $T_X:= T(X)$ through a classical-quantum (CQ) channel $T: \mathcal{X} \rightarrow \mathcal{D}(\mathcal{H}_d)$.
The leakage of private information through the output quantum state is bounded by a certain threshold if $T$ is a QLDP mechanism, as defined below \cite{zhouDifferentialPrivacyQuantum2017a, yoshidaClassicalMechanismOptimal2020, hircheQuantumDifferentialPrivacy2023}.

\begin{definition}
    For $\epsilon > 0$, an $\epsilon$-QLDP mechanism is a CQ channel $T:\mathcal{X} \rightarrow \mathcal{D}(\mathcal{H}_d)$ such that for all POVM $\Lambda = \{\Lambda_y\}_{y \in \mathcal{Y}}$ and all $x,x' \in \mathcal{X}, y \in \mathcal{Y}$,
    \begin{equation}
        \mathrm{Tr}(\Lambda_y T_x) \leq e^\epsilon \mathrm{Tr}(\Lambda_y T_{x'}).
    \end{equation}
\end{definition}

Note that $T$ is an $\epsilon$-QLDP mechanism if and only if
\begin{equation}
    T_x \leq e^\epsilon T_{x'},    
\end{equation}
for all {\revv{$x,x' \in \mathcal{X}$}}.
Here, $\epsilon>0$ can be interpreted as the maximum allowable leakage of private information through $T$.

An LDP mechanism that perturbs $X$ into a classical random variable $Y$ is an instance of a QLDP mechanism.
Specifically, an $\epsilon$-LDP mechanism is an $\epsilon$-QLDP mechanism whose output states $\{ T_x \}_{x \in \mathcal{X}}$ can be simultaneously diagonalized with respect to an orthonormal basis of $\calH_d$, that is, {\revv{$T_x = \mathrm{diag}(q_x)$}} for some probability vector $q_x \in \mathbb{R}^d$.
Consequently, if $\mathcal{X} = [v]$, a classical privacy mechanism $T$ can be identified with a row stochastic matrix $q \in \mathbb{R}^{v \times d}$ where $q_{xy}$ denotes the $y$-th component of $q_x$.
This identification recovers the original definition for an $\epsilon$-LDP mechanism \cite{kasiviswanathanWhatCanWe2011a,duchiLocalPrivacyStatistical2013a, dworkAlgorithmicFoundationsDifferential2013}.
\begin{definition}
For $\epsilon > 0$, an $\epsilon$-LDP mechanism is a row stochastic matrix $q \in \mathbb{R}^{v \times d}$ such that for all $x,x' \in [v]$ and $y\in [d]$,
    \begin{equation}
        q_{xy} \leq e^\epsilon q_{x'y}.
    \end{equation}
\end{definition}

\subsection{Hypothesis testing}

In the hypothesis testing scenarios that we consider, there are i.i.d. classical data $X_1, \ldots, X_n$ and they are perturbed into quantum states $T_{X_1} , \ldots , T_{X_n}$.
A data collector then performs a measurement on the quantum states to infer the true underlying hypothesis.
We introduce two types of hypothesis testing, symmetric and asymmetric testing.

\subsubsection{Symmetric testing}

In symmetric testing, classical data follows a distribution $P^h \in \{P^1, \ldots, P^H\}$ with prior probability $\gamma_h > 0$.
A data collector performs the POVM $\{\Lambda_h\}_{h \in [H]}$ on the quantum states $T_{X_1} , \ldots, T_{X_n}$ to infer the true hypothesis $h$.

The minimum average error probability can be written as
\begin{multline}
    P_e^{(n)}(T; \gamma,\{P^h\}_{\rev h \in [H]})
    \\ = \inf\limits_{\{\Lambda_h\}_{\rev h\in [H]}: \text{POVM}} \sum\limits_{h \neq h'} \mathrm{Tr}(\gamma_h \rho_{h}^{\otimes n} \Lambda_{h'}),
\end{multline}
where $\rho_{h} = \sum_x P^h_x T_x$ refers to the ensemble of output states $\{P^h_x , T_x \}_{x \in \mathcal{X}}$.
The minimum average error probability $P_e^{(n)}$ decays exponentially to zero at the rate given by the pairwise-minimum Chernoff information \cite{liDiscriminatingQuantumStates2016}: 
\begin{align}\label{eq:min_C_dist}
    \!\!\! S(T;\{P^h\}_{h \in {\revv{[H]}}})& \! := \! \lim\limits_{n\rightarrow\infty}-\frac{1}{n} \log P_e^{(n)}(T; \gamma,\{P^h\}_{h \in {\revv{[H]}}})
    \\& = \min\limits_{h \neq h'} C(\rho_h,\rho_{h'}),
\end{align}
where the Chernoff information $C(\rho,\sigma)$ is defined as
\begin{equation}
    C(\rho,\sigma):= -\log \min\limits_{s \in [0,1]} \mathrm{Tr}(\rho^s \sigma^{1-s}).
\end{equation}
Note that the above rate does not depend on $\gamma$, and it is called the error exponent in symmetric testing.

When classical data is perturbed by an LDP mechanism $q$, the output distribution is given by $\tilde{q}_{P^h}:= q^\top P^h$ when $X \sim P^h$.
Also, the error exponent in \eqref{eq:min_C_dist} recovers the classical result \cite{levitanCompetitiveNeymanPearsonApproach2002},
\begin{equation}
    S(q;\{P^h\}_{h \in {\revv{[H]}}}) = \min\limits_{h \neq h'} C(\tilde{q}_h,\tilde{q}_{h'}),
\end{equation}
where the Chernoff information for two probability vectors $p,r$ is given by
\begin{equation}
    C(p,r) = -\log \min\limits_{s \in [0,1]} \sum\limits_{y \in [d]} p_{y}^s r_{y}^{1-s}.
\end{equation}

\subsubsection{Asymmetric hypothesis testing}

In this scenario, we consider the following two hypotheses:
\begin{itemize}
    \item Null hypothesis: $X \sim P$ for some $P \in \mathcal{N}$,
    \item Alternative hypothesis: $X \sim P$ for some $P \in \mathcal{A}$,
\end{itemize}
where $\mathcal{N}$ and $\mathcal{A}$ are sets of distributions supported on $\mathcal{X}$.
The data collector aims to determine whether the null or alternative hypothesis holds by performing a POVM $\{\Lambda, I-\Lambda\}$ on the revealed states.
If the data collector observes the measurement outcome corresponding to $\Lambda$, it concludes that the null hypothesis holds; otherwise, the alternative hypothesis is accepted.

There are two types of error probabilities:
\begin{itemize}
    \item Type I: $\alpha^{(n)}(T,\Lambda; \mathcal{N}) := \max\limits_{P \in \mathcal{N}} \mathrm{Tr}(\rho_P^{\otimes n} (I-\Lambda) )$,
    \item Type II: $\beta^{(n)}(T,\Lambda; \mathcal{A}) := \max\limits_{P \in \mathcal{A}} \mathrm{Tr}(\rho_P^{\otimes n} \Lambda )$,
\end{itemize}
where $\rho_P = \sum_x P_x T_x$ refers to the ensemble of output states $\{P_x , T_x \}_{x \in \mathcal{X}}$.
Since there is a trade-off between the two error probabilities, the optimal measurement in asymmetric testing is typically defined as the one that minimizes the type II error probability subject to the constraint on the type I error not exceeding a threshold $\delta \in (0,1)$.
Accordingly, the optimal type II error probability is defined as
\begin{equation}
    \beta^{(n)}_\delta (T; \mathcal{N},\mathcal{A}) := \inf\limits_{\Lambda: \substack{0 \leq \Lambda \leq I, \\ \alpha^{(n)}(T,\Lambda; \mathcal{N}) \leq \delta}} \beta^{(n)} (T,\Lambda; \mathcal{A}).
\end{equation}

If $\mathcal{A}$ is a singleton $\{P^*\}$, then the previous results \cite{hayashiOptimalSequenceQuantum2002,bjelakovicQuantumVersionSanovs2005} imply that $\beta^{(n)}_\delta$ decays exponentially to zero at the rate given by the minimum relative entropy, i.e.,
\begin{align}\label{eq:min_rel_ent}
    A(T; \mathcal{N},\{P^*\}) &:=\lim\limits_{n\rightarrow\infty}-\frac{1}{n} \log \beta^{(n)}_\delta (T; \mathcal{N},\{P^*\})
    \\& = \min\limits_{P \in \mathcal{N}} D(\rho_P \| \rho_{P^*}),
\end{align}
where $D(\rho \| \sigma)$ denotes the relative entropy,
\begin{equation}
    D(\rho \| \sigma) := \mathrm{Tr}(\rho (\log \rho - \log \sigma)).
\end{equation}
Note that the above rate does not depend on $\delta$, and it is called the error exponent in asymmetric testing.

Similar to the symmetric testing scenario, when classical data is perturbed by an LDP mechanism $q$, the error exponent in \eqref{eq:min_rel_ent} recovers the classical result \cite{leangAsymptoticsMhypothesisBayesian1997},
\begin{equation}
    A(q; \mathcal{N},\{P^*\}) = \min\limits_{P \in \mathcal{N}} D(\tilde{q}_P \| \tilde{q}_{P^*}),
\end{equation}
where $\tilde{q}_P = q^\top P$, and the relative entropy for probability vectors $p,r \in \mathbb{R}^d$ is defined as
\begin{equation}
    D(p \| r) := \sum\limits_{y \in [d]} p_y \log \frac{p_y}{r_y}.
\end{equation}

\section{Problem formulation}\label{sec:prob_formulation}

{\rev 
Before presenting the formal model, we provide a practical scenario that motivates our work. Consider a survey investigating user preferences for $v$ products, where individuals may strongly prefer a single product while having uniform preferences among the remaining items.
In this context, the inference objectives of symmetric and asymmetric hypothesis testing can be distinguished as follows:
\begin{enumerate}
    \item Symmetric hypothesis testing\\
    The goal is to identify which specific product among the $v$ items is the most preferred.
    \item Asymmetric hypothesis testing\\
    The goal is to determine whether a specific product is highly preferred or preferences are uniform across all products.
\end{enumerate}
Since individual responses may reveal private information, each individual randomizes their response using a QLDP or LDP mechanism before sending it to the data collector.
We use this scenario as a primary motivating example throughout the paper.

Let us now introduce the formal definition of the scenario and the problem of quantum advantage under consideration.}
For both symmetric and asymmetric hypothesis testing, there are $n$ data providers and a single data collector.
For each $i \in [n]$, the $i$-th provider holds classical categorical data $X_i \in [v]$ for some $v \geq 2$.
We assume that $X_1,\ldots,X_n$ are i.i.d. according to a distribution $P$ supported on $[v]$.

In symmetric testing, we assume that $P$ can be one of $\{P^{h,\eta}\}_{h \in [v]}$ for a given $\eta \in (0,1]$, that we {\revv refer to as} smoothed point mass distributions, where
\begin{equation}
    P^{h,\eta}_x := \eta \delta_{h,x} + \frac{1-\eta}{v}.
\end{equation}
In asymmetric testing, we consider 
\begin{itemize}
    \item Null hypothesis: $P \in \{P^{h,\eta}\}_{h \in [v]}$,
    \item Alternative hypothesis: $P = P^0$,
\end{itemize}
where $P^0$ denotes the uniform distribution on $[v]$.

Each data provider perturbs its data into a quantum state by using an $\epsilon$-QLDP mechanism $T:[v] \rightarrow \mathcal{D}(\calH_d)$ for some $d \geq 2$, and reveals the resulting quantum state to the data collector.
{\revv We note that $v \geq 2$ is given by the scenario, and assume that $d \geq 2$ can be chosen as an arbitrary finite integer independent of $v$.}
We denote the ensemble of output states by
\begin{equation}
    \rho_{h,\eta} := \sum\limits_{x \in [v]} P_x^{h,\eta} T_x, \quad \rho_0 := \sum\limits_{x \in [v]} P_x^{0} T_x.
\end{equation}

The utilities of a QLDP mechanism $T$ in symmetric and asymmetric testing are measured by the error exponents,
\begin{align}
    S^\eta(T) &:= S(T; \{P^{h,\eta}\}_{h \in [v]}),
    \\ A^\eta(T) &:= A(T; \{P^{h,\eta}\}_{h \in [v]}, \{P^0\}),
\end{align}
respectively.
Accordingly, we define the (optimal) quantum \emph{privacy-utility trade-off (PUT)} as the maximum utility achievable by an $\epsilon$-QLDP mechanism:
\begin{align}
    S_\mathrm{Q}^\eta(v,\epsilon) &:=\sup\limits_{T: \epsilon \textnormal{-QLDP}} S^\eta(T),
    \\ A_\mathrm{Q}^\eta(v,\epsilon) &:=\sup\limits_{T: \epsilon \textnormal{-QLDP}} A^\eta(T).
\end{align}

In a fully classical scenario, each data provider perturbs its data by using an LDP mechanism $q \in \mathbb{R}^{v\times b}$ for some $b \geq 2$.
We denote the output distributions by 
\begin{equation}
    \tilde{q}_{h,\eta}:=q^\top P^{h,\eta}, \quad \tilde{q}_{0}:=q^\top P^{0}.
\end{equation}
The utilities of $q$ in symmetric and asymmetric testing are
\begin{align}
    S^\eta(q) &:= S(q; \{P^{h,\eta}\}_{h \in [v]}),
    \\ A^\eta(q) &:= A(q; \{P^{h,\eta}\}_{h \in [v]}, \{P^0\}),
\end{align}
respectively.
Accordingly, the (optimal) classical PUT is defined by
\begin{align}
    S_\mathrm{C}^\eta(v,\epsilon) &:=\sup\limits_{q: \epsilon \textnormal{-LDP}} S^\eta(q),
    \\ A_\mathrm{C}^\eta(v,\epsilon) &:=\sup\limits_{q: \epsilon \textnormal{-LDP}} A^\eta(q).
\end{align}

We say there is a \emph{quantum advantage} if the quantum PUT is strictly greater than the classical PUT, i.e., $S^\eta_{\mathrm{Q}} > S_{\mathrm{C}}^\eta$ or $A^\eta_{\mathrm{Q}} > A_{\mathrm{C}}^\eta$.

\begin{remark}
    For $v=2$, there is no quantum advantage.
    This result is a direct consequence of the previous results in \cite{yoshidaClassicalMechanismOptimal2020}.
    In detail, when $v=2$, every $\epsilon$-QLDP mechanism can be simulated by a sequential composition of an $\epsilon$-LDP mechanism and a CQ channel (that is, it is essentially classical).
    Since both the Chernoff information and the relative entropy satisfy the data-processing inequality (DPI) \cite[Chap.~4.4.1]{tomamichelQuantumInformationProcessing2016}, there is no quantum advantage when $v=2$.
\end{remark}

\begin{remark}
    In general, there is no quantum advantage with respect to the accuracy for statistical inference based on classical data when privacy constraints are absent.
    To elaborate, let $Z$ denote a random variable representing the statistical information that the data collector aims to infer, and let the collected data $X^n$ be distributed according to a distribution $P_{X^n|Z}$.
    The data collector produces an estimate $\hat{Z}$ of $Z$ from $X^n$ using quantum processing.
    Quantum processing is described by first preparing a quantum state corresponding to $X^n$, and then performing a measurement on this state to obtain $\hat{Z}$.
    Note that the accuracy of $\hat{Z}$ depends on quantum processing only through the conditional probability $P_{\hat{Z}|X^n}$ that the processing induces, and any conditional distribution $P_{\hat{Z}|X^n}$ can be simulated classically.
    Thus, there is no quantum advantage in this setup.
\end{remark}

\begin{remark}
    The comparison between quantum and classical PUTs is fair, as an $\epsilon$-QLDP mechanism provides exactly the same level of privacy protection as an $\epsilon$-LDP mechanism from an operational perspective.
    For a given $\epsilon$-LDP mechanism, $\epsilon$ captures the limit of the probability of adversarial guessing on any discrete private data based on the output of the $\epsilon$-LDP mechanism \cite[Thm.~14]{issaOperationalApproachInformation2020}.
    When a data provider uses an $\epsilon$-QLDP mechanism to perturb data, an adversary {\revv would have to} perform a measurement on the output quantum state to infer private information.
    As a direct consequence of \cite[Thm.~14]{issaOperationalApproachInformation2020} and the definition of $\epsilon$-QLDP, $\epsilon$ also captures the limit of the probability of adversarial guessing on private information for all possible measurements on the output quantum state of a given $\epsilon$-QLDP mechanism.
\end{remark}

\section{Summary of main results}\label{sec:main}

The main contribution of our work is to prove a quantum advantage in both private symmetric and asymmetric hypothesis testing scenarios for a set of parameters specified below.
\begin{theorem}\label{thm:main}
    If $v=4$ or $9$, then there exist $\epsilon_0 >0$ and $\eta_0 < 1$ such that for all $\epsilon \in (0,\epsilon_0]$ and $\eta \in [\eta_0,1]$,
    \begin{equation}
        S_{\mathrm{Q}}^\eta (v,\epsilon) > S_{\mathrm{C}}^\eta (v,\epsilon), \; A_{\mathrm{Q}}^\eta (v,\epsilon) > A_{\mathrm{C}}^\eta (v,\epsilon).
    \end{equation}
    Additionally, if $\eta = 1$, then for all $3 \leq v \leq 9$, there exists $\epsilon_1 >0$ such that for all $\epsilon \in (0,\epsilon_1]$,
    \begin{equation}
        \quad S^1_{\mathrm{Q}}(v,\epsilon) > S^1_{\mathrm{C}}(v,\epsilon).
    \end{equation}
\end{theorem}

We prove the theorem by first proposing a specific $\epsilon$-QLDP mechanism $T^*$ (Section~\ref{sec:Q_mech}), and deriving an upper bound $\overline{S^\eta_{\mathrm{C}}}$ on $S^\eta_\mathrm{C}$ and characterizing $A^\eta_\mathrm{C}$ (Section~\ref{sec:CPUT}).
Here, $\overline{S^1_{\mathrm{C}}} = S^1_{\mathrm{C}}$ also holds.
Next, we show that at $\eta = 1$ and for the values of $v$ specified in the theorem, both $S^1(T^*) / S^1_{\mathrm{C}}$ and $A^1(T^*) / A^1_{\mathrm{C}}$ converge to {\rev values greater than $1$} as $\epsilon$ approaches zero (Section~\ref{sec:Q_adv_pf}).
This implies the theorem by the continuity in $\epsilon$ and $\eta$.
We note that our results concerning $S^\eta_\mathrm{C}$ and {\revv {$A^\eta_\mathrm{C}$}} also constitute a contribution in their own right to the literature on classical private statistical inference.

\begin{figure}[!ht]
     \centering
     \begin{subfigure}[b]{0.4\textwidth}
         \centering
         \includegraphics[width=\textwidth]{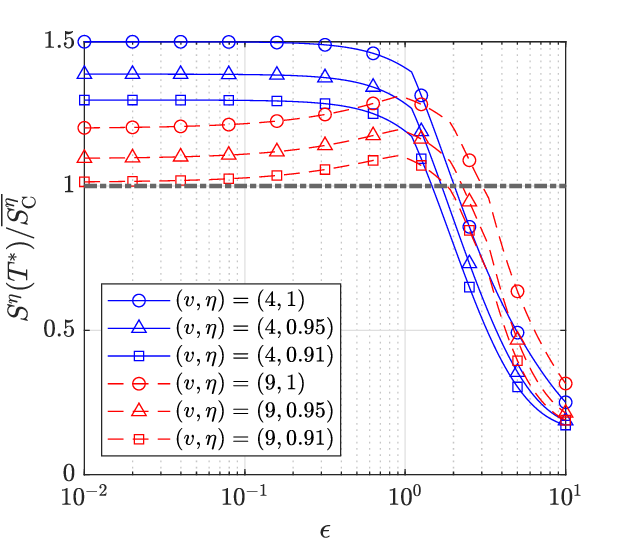}
         \caption{Symmetric testing}
     \end{subfigure}
     \begin{subfigure}[b]{0.4\textwidth}
         \centering
         \includegraphics[width=\textwidth]{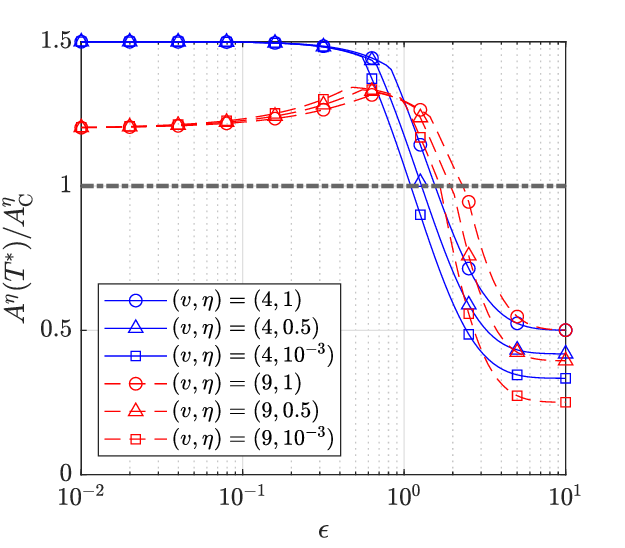}
         \caption{Asymmetric testing}
     \end{subfigure}
     \caption{Quantum advantage in private hypothesis testing}
     \label{fig:main}
\end{figure}

In Fig.~\ref{fig:main}, we plot the ratios 
\bea
\frac{ S^\eta(T^*) }{  \overline{S^\eta_{\mathrm{C}}}}~~\mathrm{and} ~~
\frac{ A^\eta(T^*) }{ A^\eta_{\mathrm{C}}}\nonumber 
\eea
for $v=4$ and $9$.
There is a quantum advantage if such ratios are greater than $1$.
As depicted in the figure, quantum advantage exists in certain parameter regimes where the curves for these ratios exceed the black horizontal line at $1$.
Although Theorem~\ref{thm:main} does not specify the values of $\epsilon_0, \eta_0$ and $\epsilon_1$, the numerical results show that these parameters are not negligible: $\epsilon_0, \epsilon_1 \geq 1$ and $\eta_0 \leq 0.91$.
Moreover, when restricted to asymmetric testing, $\eta_0$ can be approached as close as $10^{-3}$.

{\rev The above results imply that the proposed QLDP mechanism outperforms classical LDP mechanisms in the regime of relatively small $\epsilon$ for both symmetric and asymmetric testing, and $\eta$ close to $1$ for symmetric testing.
The regime of small $\epsilon$ is particularly relevant to applications involving sensitive data such as healthcare, finance, or census data, where stringent privacy protection is required.
The regime where $\eta$ is close to $1$ corresponds to the user preference survey scenario introduced in Section~\ref{sec:prob_formulation}, where there is prior knowledge that a specific item has a much higher preference than the other items.
}

\section{Proposed QLDP mechanism} \label{sec:Q_mech}

To demonstrate quantum advantage, we design a QLDP mechanism that balances the trade-off between privacy leakage and utility more effectively than the optimal LDP mechanism.
Intuitively, QLDP and utilities in both symmetric and asymmetric testing scenarios depend on the distinguishability of output states $\{T_x\}_{x \in [v]}$.
QLDP directly quantifies this distinguishability, while the utilities reflect it indirectly through the distinguishability of the ensemble states ($\{\rho_{h,\eta}\}_{h \in[v]}$ in symmetric testing, and $\{\rho_{h,\eta}, \rho_0\}$ for each $h \in [v]$ in asymmetric testing).

In balancing these two notions of distinguishability, a quantum system provides more {\rev degrees} of freedom in designing privacy mechanisms: output states can be prepared in distinct orthonormal bases for each input $x \in [v]$, whereas classical privacy mechanisms prepare all output states in the same orthonormal basis.
Specifically, a QLDP mechanism can be specified through spectral decomposition as
\begin{equation}
    T_x = \sum\limits_{i \in [d]} p_{x,i} \kb{\psi_{x,i}},
\end{equation}
where, for all $x \in [v]$, $(p_{x,i})_{i \in [d]}$ is a probability vector and $\{\ket{\psi_{x,i}}\}_{i \in [d]}$ is an orthonormal basis of $\mathcal{H}_d$.
In general, $\{\ket{\psi_{x,i}}\}_{i \in [d]}$ depends on $x$, whereas it is independent of $x$ for an LDP mechanism.
The use of distinct bases for different inputs affects the distinguishability of output states, as it influences the degree of orthogonality, i.e., the inner product between the output states.

With this idea, we consider a QLDP mechanism that first prepares {\revv a pure quantum state chosen from a set of mutually non-orthogonal states,} and then applies depolarizing noise.
Specifically, we consider a mechanism $T$ of the following form:
\begin{equation}\label{eq:achiev_scheme}
    \forall x \in [v], \quad T_x = \frac{\mu}{d} I_d + (1-\mu) \kb{\psi_x},
\end{equation}
where each $\ket{\psi_x} \in \calH_d$ is a unit vector and $\mu \in [0,\frac{d}{d-1}]$.\footnote{The depolarizing channel with parameter $\mu$ is a valid quantum channel (trace-preserving and completely positive) if and only if $\mu \in [0, \frac{d^2}{d^2-1}]$.
Though, we allow $\frac{d^2}{d^2-1} < \mu \leq \frac{d}{d-1}$ since $T$ remains a valid classical-quantum channel if and only if $\mu \in [0,\frac{d}{d-1}]$.}
If the states {\revv{$\{\ket{\psi_x}\}_{x\in[v]}$} are mutually orthonormal, in particular when $v = d$ and they form an orthonormal basis,} this mechanism reduces to the well-known LDP mechanism, called the randomized response \cite{warnerRandomizedResponseSurvey1965,kairouzExtremalMechanismsLocal2016b}, which subjects its input to uniform random noise.
Beyond this case, the trade-off between privacy leakage and utility can be balanced by jointly adjusting the depolarizing parameter $\mu$ and the degree of non-orthogonality of {\revv{$\{\ket{\psi_x}\}_{x\in[v]}$}}.
This class of mechanisms of the form in \eqref{eq:achiev_scheme} with $\mu \in (0,1]$ was previously considered in \cite{yoshidaMathematicalComparisonClassical2025}.
Compared to previous works that considered a depolarizing channel as a QDP or QLDP mechanism \cite{zhouDifferentialPrivacyQuantum2017a,angrisaniUnifyingFrameworkDifferentially2023,farokhiQuantumPrivacyHypothesisTesting2023,hircheQuantumDifferentialPrivacy2023,nuradhaQuantumPufferfishPrivacy2024,guanOptimalMechanismsQuantum2024,nuradhaContractionPrivateQuantum2025,yoshidaMathematicalComparisonClassical2025}, we further choose the pure states $\{\ket{\psi_x}\}_{x \in [v]}$ to form a subset of symmetric informationally complete (SIC) states and partially justify this choice with Proposition~\ref{prop:opt_achiev_utility}.

We remark that quantumness in the proposed QLDP mechanism lies in a collection of pure states for the manipulation of indistinguishability.
Both LDP and QLDP exploit noise to manipulate indistinguishability for optimal PUT, on top of which a QLDP mechanism applies non-orthogonal states with intrinsic indistinguishability.
We recall that quantum indistinguishability cannot be demonstrated in a classical model \cite{PhysRevX.8.011015, PRXQuantum.3.030337} and is strictly constrained by physical principles such as the relativistic causality \cite{PhysRevLett.107.170403}.
Note also that for a task of state discrimination, at most $d^2$ POVM elements in a $d$-dimensional Hilbert space generally suffice to manipulate indistinguishability \cite{10.1109/TIT.1978.1055941}.

In what follows, we present a QLDP mechanism that outperforms LDP ones by exploiting quantumness resulting from the intrinsic indistinguishability of non-orthogonal quantum states.
We focus on a QLDP mechanism in \eqref{eq:achiev_scheme} and seek to find the optimal value of $\mu$ and the pure states $\{\ket{\psi_x}\}_{x\in[v]}$ that maximize the utility while satisfying the $\epsilon$-QLDP constraint.

\subsection{Depolarizing parameter to satisfy $\epsilon$-QLDP}
In the following proposition, we derive a necessary and sufficient condition for such a privacy mechanism to be an $\epsilon$-QLDP mechanism.
Its proof, which is provided in Section~\ref{sec:Q_achiev_QLDP_pf}, is similar to that of \cite[Lem.~5.5]{yoshidaMathematicalComparisonClassical2025}.
\begin{proposition}\label{prop:achiev_QLDP}
    For any given unit vectors $\{\ket{\psi_x}\}_{x \in [v]} \subset \calH_d$, a privacy mechanism $T$ of the form in \eqref{eq:achiev_scheme} is an $\epsilon$-QLDP mechanism if and only if
    \begin{equation}\label{eq:QLDP_mu}
        \mu \in \left[ \frac{dg^-_*}{dg^-_*-1}, \frac{dg^+_*}{dg^+_*-1} \right],
    \end{equation}
    where
    \begin{equation}
        c_{x,x'} := |\inp{\psi_x}{\psi_{x'}}|^2,\quad c_* := \min_{x\neq x'} c_{x,x'},
    \end{equation}
    and
    \begin{equation}\label{eq:g_star}
        \quad g^{\pm}_* := \frac{1 \pm \sqrt{1+\frac{1-c_*}{\sinh^2(\epsilon/2)}}}{2}.
    \end{equation}
\end{proposition}

Note that the term $1-c_*$ in \eqref{eq:g_star} represents the maximum cosine distance between the pure states $\{\ket{\psi_x}\}_{x \in [v]}$.
The above proposition implies that the minimum amount of depolarizing noise required to satisfy $\epsilon$-QLDP constraint increases as the pure states $\{\ket{\psi_x}\}_{x \in [v]}$ become more separated and as $\epsilon$ decreases.

It can be intuitively expected that, for a given $\{\ket{\psi_x}\}_{x \in [v]}$, the minimum $\mu$ in \eqref{eq:QLDP_mu} maximizes the utility from the perspective of the DPI.
A non-trivial challenge lies in choosing the set of pure states $\{\ket{\psi_x}\}_{x \in [v]}$.

\subsection{Choice of pure states}

In symmetric testing, the utility is defined in terms of the Chernoff information, which can be interpreted as a measure of distance between states.
Consequently, the utility may be maximized when the minimum distance among the states $\{\ket{\psi_x}\}_{x \in [v]}$ is maximized.
One candidate set that achieves this max-min distance is a set of uniformly separated states.

In contrast, Proposition~\ref{prop:achiev_QLDP} indicates that a greater amount of depolarizing noise is required to satisfy the $\epsilon$-QLDP constraint as the maximum distance (with respect to the cosine distance) between the states increases.
Since increasing depolarizing noise reduces the utility, it is not evident that states achieving the max-min distance maximize the utility.
At this point, the following proposition shows that, at $\eta = 1$ and for $\epsilon \approx 0$, the utility in symmetric testing is maximized when the pure states are chosen to be uniformly separated in a given dimension.
Its proof is in Section~\ref{sec:Q_achiev_pf}.

\begin{proposition}\label{prop:opt_achiev_utility}
    For any given $\{\ket{\psi_x}\}_{x\in [v]} \subset \calH_d$, and a privacy mechanism $T$ of the form in \eqref{eq:achiev_scheme}, we have
    \begin{align}
        \max\limits_{\mu}S^1(T) &= -\log G(c^*,d,\mu_*) \label{eq:Q_PUT_eq}
        \\& \overset{\epsilon\rightarrow 0}{\approx} \frac{1- c^*}{4d(1- c_*)}\epsilon^2 \leq \frac{\epsilon^2}{4d}, \label{eq:Q_PUT_approx}
    \end{align}
    where the maximization over $\mu$ is subject to the bounds in \eqref{eq:QLDP_mu}, 
    \begin{multline}\label{eq:large_G}
        G(c,d,\mu)
        \\ := c + \frac{1-c}{d} \left((d-2)\mu + 2 \sqrt{\mu(d-(d-1)\mu)} \right),
    \end{multline}
    \begin{equation}\label{eq:mu_star}
        c^* := \max_{x \neq x'} c_{x,x'}, \quad \mu_* := \frac{dg_*^-}{dg_*^- -1},
    \end{equation}
    and $c_*, g_*^-$ are defined in \eqref{eq:g_star}.
    The last inequality becomes an equality if and only if $c_{x,x'}$ is constant for all $x \neq x'$.
\end{proposition}

According to the above proposition, we narrow our focus to the set of pure states $\{ \ket{\psi_x} \}_{x \in [v]}$ such that $|\inp{\psi_x}{\psi_{x'}}|^2$ is constant for all $x \neq x'$ (i.e., uniformly separated).
Then, the utility in symmetric testing at $\eta=1$ and $\epsilon \approx 0$ is maximized when such $\{ \ket{\psi_x} \}_{x \in [v]}$ lie in the minimum possible dimension, as in \eqref{eq:Q_PUT_approx}.
It is known that this minimum dimension is $\left\lceil \sqrt{v} \right\rceil$ \cite{zaunerQuantumDesignsFoundations2011}, and a subset of SIC states in dimension $\left\lceil \sqrt{v} \right\rceil$ satisfies all the desired conditions.\footnote{Originally, SIC states are defined via a SIC-POVM, which {\revv is} a specific instance of {\rev POVM.}
A SIC-POVM can equivalently be characterized as a set of pure states, as defined in Definition~\ref{def:SIC_POVM} \cite{renesSymmetricInformationallyComplete2004,zaunerQuantumDesignsFoundations2011}.
In this paper, we use the term “SIC states” because we utilize this concept with its physical meaning as quantum states.}

\begin{definition}\label{def:SIC_POVM}
    A set of SIC states is a set of $d^2$ unit vectors $\{\ket{\psi_x}\}_{x\in[d^2]}$ in $\calH_d$ satisfying
    \begin{equation}\label{eq:SIC_POVM_res_id}
        \sum\limits_{x=1}^{d^2} \kb{\psi_x} = d I_d,
    \end{equation}
    and
    \begin{equation}
        \forall x \neq x', \quad |\inp{\psi_x}{\psi_{x'}}|^2 = \frac{1}{d+1}.
    \end{equation}
\end{definition}

\subsection{Proposed QLDP mechanism}\label{sec:prop_mech}
Based on the arguments developed in this section, we propose the QLDP mechanism $T^*$, which takes the form given in \eqref{eq:achiev_scheme}, where $\{\ket{\psi_x}\}_{x\in[v]}$ is chosen as a subset of SIC states in dimension ${\left\lceil \sqrt{v} \right\rceil}$, and $\mu$ is set to $\mu_*$ defined in \eqref{eq:mu_star}.
For certain parameter regimes, we derive closed-form expressions for the utilities of $T^*$ in symmetric and asymmetric testing in the following proposition, whose proof is provided in Section~\ref{sec:Q_achiev_proposed_pf}.

\begin{proposition}\label{prop:utility_cl_form}
    For the proposed QLDP mechanism $T^*$, we have 
    \begin{align}
        S^1(T^*) &= -\log G\left(\frac{1}{\left\lceil \sqrt{v} \right\rceil+1}, \left\lceil \sqrt{v} \right\rceil, \mu_* \right), \label{eq:S_1}
    \end{align}
    where $G$ is defined in \eqref{eq:large_G}, and $\mu_*$ in \eqref{eq:mu_star} is calculated by substituting $c_*=1/(\left\lceil \sqrt{v} \right\rceil+1)$.
    Moreover, if $v=d^2$ for some integer $d \geq 2$, then
    \begin{equation}\label{eq:S_eta}
        S^\eta(T^*) = -\log G\left(\frac{1}{d+1}, d, \mu_\eta \right),
    \end{equation}
    \begin{equation}\label{eq:A_eta}
        A^{\eta}(T^*) = \log d + L\left( 1-\mu_\eta + \frac{\mu_\eta}{d}\right) + (d-1)L\left(\frac{\mu_\eta}{d}\right),
    \end{equation}
    where
    \begin{equation}
        \mu_\eta := 1-\eta + \mu_* \eta, \quad L(x) := x\log x.
    \end{equation}
\end{proposition}

Although the choice of pure states is partially justified only in the symmetric testing scenario with $\eta=1$ and $\epsilon \approx 0$, our QLDP mechanism $T^*$ demonstrates a quantum advantage in both symmetric and asymmetric testing, {\rev as stated in Theorem~\ref{thm:main}.
Moreover, the numerical evaluation in Fig.~\ref{fig:main} shows that quantum advantage persists for $\epsilon \leq 1$, with $\eta \geq 0.91$ for symmetric testing and $\eta \geq 10^{-3}$ for asymmetric testing.}

\section{Classical PUT}\label{sec:CPUT}

In this section, we characterize an upper bound $\overline{S^\eta_\mathrm{C}}$ on the classical PUT $S^\eta_\mathrm{C}$ in symmetric testing, as well as the exact classical PUT $A^{\eta}_\mathrm{C}$ in asymmetric testing.
We note that the upper bound $\overline{S^\eta_\mathrm{C}}$ is tight when $\eta = 1$.

\begin{proposition}\label{prop:CPUT}
    For all $v \geq 2, \epsilon > 0$, and $\eta \in (0,1]$, the classical PUT in a symmetric testing scenario is upper bounded as
    \begin{align}
        &S^\eta_\mathrm{C}(v,\epsilon) \leq \overline{S^\eta_{\mathrm{C}}}(v,\epsilon) \label{eq:CPUT_S_UB}
        \\& = -\log \left(1 - \frac{(v+\eta^2 -1)(e^{\epsilon/2}-1)^2}{v^2(v-1)} \max\limits_{k \in [0:v]} \frac{k(v-k)}{f(v,k,\epsilon)} \right). \nonumber
    \end{align}
    Moreover, $S^1_\mathrm{C}(v,\epsilon) = \overline{S^1_{\mathrm{C}}}(v,\epsilon)$.
    In asymmetric testing scenario, the classical PUT is characterized by
    \begin{equation}
        A^\eta_\mathrm{C}(v,\epsilon) = \max\limits_{k\in [0:v]} \frac{F(v,k,\epsilon)}{v f(v,k,\epsilon)}, \label{eq:CPUT_A_UB}
    \end{equation}
    where    
    \begin{align}
        f(v,k,\epsilon) &:= \frac{ke^\epsilon + v-k}{v}, \label{eq:not_A_start}
        \\ \Delta_1 &:= {\rev \eta  e^\epsilon + (1-\eta) f(v,k,\epsilon)},
        \\ \Delta_2 &:= {\rev \eta  + (1-\eta) f(v,k,\epsilon)},
    \end{align}
    \begin{equation}
        F(v,k,\epsilon) := kL (\Delta_1) +(v-k)L(\Delta_2) -v L(f(v,k,\epsilon)), \label{eq:not_A_end}
    \end{equation}
    and $L(x) = x\log x$.
\end{proposition}

We first derive upper bounds on the classical PUTs (converse part), and then show that a block design mechanism \cite{parkExactlyOptimalCommunicationEfficient2024} achieves the upper bound $\overline{S^\eta_\mathrm{C}}$ at $\eta=1$, and $A^\eta_\mathrm{C}$ for all $\eta \in (0,1]$ (achievability part).

\subsection{Converse: Classical PUT} \label{sec:CPUT_conv}

In both symmetric and asymmetric testing scenarios, the main idea for deriving upper bounds on the classical PUTs is to exploit the DPI.
Since both the Chernoff information and the relative entropy satisfy the DPI, we can narrow down the set of possible candidates for the optimal LDP mechanism.
Specifically, we show that every LDP mechanism can be simulated by applying post-processing to an extremal LDP mechanism \cite{kairouzExtremalMechanismsLocal2016b}.
The formal descriptions are as follows.

\begin{definition}\label{def:DPI}
    A real-valued function $U$ defined on the set of all row stochastic matrices $q$ is said to satisfy the DPI if $U(q) \geq U(q \Phi)$ for all row stochastic matrices $\Phi$ of compatible dimension.
\end{definition}

\begin{definition}\label{def:extremal_mech}
    For an integer $v \geq 2$, let
    \begin{equation}
        S^{(v)} := \mathbbm{1}_{v \times 2^v} + (e^\epsilon-1) \mathrm{BIN}^{(v)},
    \end{equation}
    where $\mathrm{BIN}^{(v)} \in \{0,1\}^{v \times 2^v}$ is the matrix whose $y$-th column corresponds to the binary representation of $y-1$.
    A row stochastic matrix $q \in \mathbb{R}^{v\times {2^v}}$ is called an extremal $\epsilon$-LDP mechanism if there exists a non-negative vector $\theta \in \mathbb{R}^{2^v}$ such that $q = S^{(v)} \mathrm{diag}(\theta)$ and $S^{(v)} \theta = \mathbbm{1}_{v}$.
\end{definition}

\begin{lemma}\label{lem:DPI_extremal}
    For any given $\epsilon$-LDP mechanism $q$, there exists an extremal $\epsilon$-LDP mechanism $q^*$ and a row stochastic matrix $\Phi$ such that $q= q^* \Phi$.
\end{lemma}
\begin{IEEEproof}
    For a given $\epsilon$-LDP mechanism $q$, we have $q = \bar{q} \cdot \mathrm{diag}(\theta)$, where {\revv{$\theta \in [0,1]^b$}}, $\theta_y = \min_{x \in [v]} q_{xy}$, and $\bar{q} \in {\rev [1,e^\epsilon]}^{v \times b}$.
    Let $\bar{q}^y$ be the $y$-th column of $\bar{q}$.
    Then, $\bar{q}^y$ is a convex combination of the columns of $S^{(v)}$, i.e., there exists $\bar{\theta}^y \in \mathbb[0,1]^{2^v}$ such that $\bar{q}^y = S^{(v)}\bar{\theta}^y$ and $\sum_{x\in 2^v} \bar{\theta}^y_x = 1$ \cite[Lem. 12]{yeOptimalSchemesDiscrete2018}.
    Thus, 
    \begin{equation}
        q = S^{(v)} \begin{bmatrix} \bar{\theta}^1, \cdots ,\bar{\theta}^b \end{bmatrix} \mathrm{diag}(\theta) = S^{(v)} \mathrm{diag}(\zeta) \bar{\zeta},
    \end{equation}
    where $\zeta \in [0,1]^{2^v}$, $\zeta_z = \sum_{y \in [b]} \theta_y \bar{\theta}^y_z$, and $\bar{\zeta}_{zy} = \theta_y \bar{\theta}^y_z / \zeta_z$.
    Clearly, $\bar{\zeta}$ is a row stochastic matrix.
    In addition, $q^* = S^{(v)} \mathrm{diag}(\zeta)$ is an extremal $\epsilon$-LDP mechanism because
    \begin{align}
        S^{(v)} \zeta &= S^{(v)} \sum\limits_{y \in [b]} \theta_y \bar{\theta}^y = \sum\limits_{y \in [b]} \theta_y S^{(v)} \bar{\theta}^y
        \\& = \sum\limits_{y \in [b]} \theta_y \bar{q}^y = {\rev{\bar{q} \theta}} = \mathbbm{1}_v. 
    \end{align}
\end{IEEEproof}

By Lemma~\ref{lem:DPI_extremal} and the DPI, we can analyze the classical PUT by maximizing the utility only over extremal mechanisms.
From this starting point, we prove the converse part of Proposition~\ref{prop:CPUT} in Section~\ref{sec:CPUT_conv_pf}.

\subsection{Achievability: Classical PUT}

For the achievability part, we show that a block design mechanism \cite{parkExactlyOptimalCommunicationEfficient2024} achieves the classical PUTs in Proposition~\ref{prop:CPUT} for $\eta = 1$ in symmetric testing, and for all $\eta \in(0,1]$ in asymmetric testing.
A block design mechanism is constructed from a balanced incomplete block design (BIBD) \cite{stinsonCombinatorialDesignsConstructions2007,colbournHandbookCombinatorialDesigns2006}.

\begin{definition}\label{def:BIBD}
    A BIBD is a pair $(V,B)$ of a finite set of vertices $V$ and a non-empty set of blocks $B \subset 2^{V}$ satisfying the following symmetries for some non-negative integers $r,k,\lambda$:
    \begin{enumerate}
        \item $r$-regular: For all $p \in V$, $|\{e \in B : p \in e\}| = r$,
        \item $k$-uniform: For all $e \in B$, $|e|= k$,
        \item $\lambda$-pairwise balanced: For all two distinct vertices $p,p'$, we have $|\{e \in B : \{p,p'\} \subset e \}| = \lambda$.
    \end{enumerate}
    We denote $v:=|V|$ and $b:=|B|$.
\end{definition}

An $r$-regular, $k$-uniform, $\lambda$-pairwise balanced BIBD is conventionally written as $(v,k,\lambda)$-design, since $r$ and $b$ are determined by $(v,k,\lambda)$.
Whenever $k\notin \{0,1,v\}$, the following identities {\revv must} hold \cite[Thm. 1.8, 1.9]{stinsonCombinatorialDesignsConstructions2007}:
\begin{equation}\label{eq:BD_necc}
    r = \frac{\lambda(v-1)}{k-1}, \quad b = \frac{vr}{k}.
\end{equation}
If $k=0,1$, and $v$, then we must have $(r,b)=(0,1),(1,v)$, and $(1,1)$, respectively.
Also, for any $k \in [v]$, there exists a $(v,k,\lambda)$-design for some $\lambda$ (the complete $k$-uniform hypergraph).

\begin{definition}\label{def:BD_mech}
    Let $(V,B) = ([v],\{e_1,\ldots,e_b\})$ be a $(v,k,\lambda)$-block design.
    A $(v,k,\lambda,\epsilon)$-block design mechanism constructed from $(V,B)$ is a row stochastic matrix $q \in \mathbb{R}^{v\times b}$ such that 
    \begin{equation}
        q_{xy} = \begin{cases}
            \frac{e^\epsilon}{re^\epsilon + b-r} & \textnormal{if }x \in e_y
            \\ \frac{1}{re^\epsilon + b-r} & \textnormal{if }x \not\in e_y
        \end{cases}.
    \end{equation}
\end{definition}

Clearly, $(v,k,\lambda,\epsilon)$-block design mechanism is an $\epsilon$-LDP mechanism.
We note that a $(v,k,\lambda,\epsilon)$-block design mechanism constructed from the complete $k$-uniform hypergraph is known as the subset selection mechanism \cite{yeOptimalSchemesDiscrete2018}.

We prove the achievability part of Proposition~\ref{prop:CPUT} in Section~\ref{sec:CPUT_achieve_pf} by calculating the utilities of a block design mechanism.

\section{Analytic proof of quantum advantage}\label{sec:Q_adv_pf}
Although we have formulas for the utilities of our proposed QLDP mechanism and the classical PUTs in Propositions~\ref{prop:opt_achiev_utility} and~\ref{prop:utility_cl_form}, it is cumbersome to directly calculate and compare them analytically by hand.
However, when $\epsilon \approx 0$ and $\eta = 1$, these formulas can be further simplified, allowing us to prove quantum advantage as stated in the following corollary.
This corollary then directly implies Theorem~\ref{thm:main} by continuity.
\begin{corollary}
    Let $T^*$ be the proposed QLDP mechanism introduced in Section~\ref{sec:prop_mech}.
    For all $v \geq 2$, we have 
    \begin{equation}
        \lim\limits_{\epsilon \rightarrow 0}\frac{S^1 (T^*)}{S^1_{\mathrm{C}}(v,\epsilon)} \geq
        \begin{cases} \frac{4(v-1)}{v\lceil \sqrt{v} \rceil} & \rev{\textnormal{if }v\textnormal{ is even}}
            \\ \frac{4v}{(v+1) \lceil \sqrt{v} \rceil} & \rev{\textnormal{if }v\textnormal{ is odd}}
        \end{cases}.
    \end{equation}
    Hence, if $3 \leq v \leq 9$, then
    \begin{equation}
         \lim\limits_{\epsilon \rightarrow 0}\frac{S^1 (T^*)}{S^1_{\mathrm{C}}(v,\epsilon)} > 1.
    \end{equation}
    Moreover, when $v=d^2$ for some integer $d \geq 2$, we have
    \begin{equation}
        \lim\limits_{\epsilon \rightarrow 0}\frac{A^1 (T^*)}{A^1_{\mathrm{C}}(v,\epsilon)} \geq \begin{cases} \frac{4(d^2-1)}{d^3} & \rev{\textnormal{if }v\textnormal{ is even}}
            \\ \frac{4d}{d^2+1} &\rev{\textnormal{if }v\textnormal{ is odd}}
        \end{cases}.
    \end{equation}
    Thus, if $v = 4$ or $9$, then
    \begin{equation}
         \lim\limits_{\epsilon \rightarrow 0}\frac{A^1 (T^*)}{A^1_{\mathrm{C}}(v,\epsilon)} > 1.
    \end{equation}
\end{corollary}
\begin{IEEEproof}
    For the symmetric testing scenario, Proposition~\ref{prop:opt_achiev_utility} implies
    \begin{equation}\label{eq:S_1_approx_Q}
        S^1(T^*) \overset{\epsilon\rightarrow 0}{\approx} \frac{\epsilon^2}{4 \lceil \sqrt{v} \rceil}.
    \end{equation}
    To approximate $S^1_\mathrm{C}(v,\epsilon)$, let
    \begin{equation}\label{eq:S_1_C_max}
        S^1_\mathrm{C}(v,\epsilon) = -\log \left(1 - \frac{(e^{\epsilon/2}-1)^2}{v-1} \max\limits_{k \in [0:v]} K_S(k)  \right),
    \end{equation}
    where
    \begin{equation}
        K_S(k):= \frac{k(v-k)}{ke^\epsilon + v-k}.
    \end{equation}
    By taking the derivative in $k$,
    \begin{equation}
        K_S'(k) = \frac{-(e^\epsilon-1)k^2 -2vk +v^2}{(k e^\epsilon + v-k)^2}.
    \end{equation}
    Since the numerator is concave in $k$, $K_S'(0) > 0$ and $K_S'(v) < 0$, $K_S'(k)=0$ has a unique solution $v/(e^{\epsilon/2}+1)$ in $k \in [0,v]$.
    Hence, one of the closest integers to $v/(e^{\epsilon/2}+1)$ becomes the maximizer $k_S^*$ in \eqref{eq:S_1_C_max}.
    For $\epsilon \approx 0$, $k_S^* \in \{ v/2, v/2 - 1 \}$ if $v$ is even, and $k_S^* \in \{ (v-1)/2, (v+1)/2 \}$ if $v$ is odd.
    
    {\rev     
    Now, we approximate $S^1_\mathrm{C}(v,\epsilon)$ using a Taylor expansion around $\epsilon = 0$.
    Straightforward algebraic manipulation yields
    \begin{equation}\label{eq:K_S_approx}
        \frac{(e^{\epsilon/2}-1)^2}{v-1} K_S(k) = \frac{k(v-k)}{4v(v-1)}\epsilon^2 + O(\epsilon^3).
    \end{equation}
    Comparing the values of the above for all candidates for $k_S^*$, and using $-\log (1-x) = x + O(x^2)$, we obtain
    \begin{equation}\label{eq:S_1_approx_C}
        S_C^1(v,\epsilon) \overset{\epsilon\rightarrow 0}{\approx} \begin{cases}
        \frac{v}{16(v-1)}\epsilon^2 &\rev{\textnormal{if }v\textnormal{ is even}}
        \\ \frac{v+1}{16v}\epsilon^2 &\rev{\textnormal{if }v\textnormal{ is odd}}
        \end{cases}.
    \end{equation}
    The first part of the corollary is proved by combining \eqref{eq:S_1_approx_Q} and \eqref{eq:S_1_approx_C}.}
    
    The proof for the asymmetric testing scenario follows almost identical steps described above.
    {\rev To approximate $A^1(T^*)$, it suffices to consider the Taylor approximation of $L(x)$ around $x = 1/d$, since $\lim_{\epsilon\rightarrow 0}\mu_* = 1$, as shown in \eqref{eq:mu_star_approx} in the proof of Proposition~\ref{prop:opt_achiev_utility}.
    Explicitly, we have
    \begin{align}
        L(x) &= -\frac{\log d}{d} + (1- \log d) \left( x - \frac{1}{d}\right)
        \\&\quad\quad + \frac{d}{2} \left(x - \frac{1}{d}\right)^2 + O \left(\left(x - \frac{1}{d}\right)^3\right). \nonumber
    \end{align}
    Combining \eqref{eq:mu_star_approx}, \eqref{eq:A_eta}, and the above expansion, we obtain
    \begin{equation}\label{eq:A_1_approx_Q}
        A^1 (T^*) \overset{\epsilon\rightarrow 0}{\approx} \frac{(d^2-1)\epsilon^2}{2d^3}.
    \end{equation}}
    
    To approximate $A^1_\mathrm{C}(v,\epsilon)$, note that
    \begin{equation}\label{eq:A_1_C_max}
        A^1_\mathrm{C}(v,\epsilon) = \max\limits_{k \in [0:v]} K_A(k),
    \end{equation}
    where
    \begin{equation}
        K_A(k) := \frac{k\epsilon e^\epsilon}{ke^\epsilon + v-k} + \log \frac{v}{ke^\epsilon + v-k}.
    \end{equation}
    By taking the derivative in $k$,
    \begin{equation}
        K_A'(k) = \frac{-(e^\epsilon-1)^2 k+ v(\epsilon e^\epsilon - e^\epsilon + 1)}{(ke^\epsilon + v-k)^2}.
    \end{equation}
    Since the numerator is a linear function in $k$, the maximizer $k_A^*$ in \eqref{eq:A_1_C_max} is one of the closest integers to
    \begin{equation}
        \frac{v (\epsilon e^\epsilon - e^\epsilon + 1)}{(e^\epsilon-1)^2}.
    \end{equation}
    Hence, for $\epsilon \approx 0$, $k_A^* \in \{ v/2, v/2 - 1 \}$ if $v$ is even, and $k_A^* \in \{ (v-1)/2, (v+1)/2 \}$ if $v$ is odd.
    
    {\rev
    Now, we approximate $K_A(k)$ around $\epsilon = 0$ using Taylor expansion, which yields
    \begin{equation}
        K_A(k) = \frac{k(v-k)}{2v^2}\epsilon^2 + O(\epsilon^3).
    \end{equation}
    Comparing the values of the above for all candidates for $k_A^*$, we get
    \begin{equation}\label{eq:A_1_approx_C}
        A_C^1(v,\epsilon) \overset{\epsilon\rightarrow 0}{\approx} \begin{cases}
        \epsilon^2/8 &\rev{\textnormal{if }v\textnormal{ is even}}
        \\ \frac{v^2-1}{8v^2}\epsilon^2 &\rev{\textnormal{if }v\textnormal{ is odd}}
        \end{cases}.
    \end{equation}
    Finally, we obtain the desired result by combining \eqref{eq:A_1_approx_Q} and \eqref{eq:A_1_approx_C}.}
\end{IEEEproof}

\section{Proofs of Propositions}\label{sec:pfs}
\subsection{Proof of Proposition~\ref{prop:achiev_QLDP}}\label{sec:Q_achiev_QLDP_pf}

    By definitions, $T$ is an $\epsilon$-QLDP mechanism if and only if for all $x,x' \in [v]$,
    \begin{align}
        & e^\epsilon  T_x \geq T_{x'}
        \\ &\Leftrightarrow \frac{\mu(e^\epsilon-1)}{d}I_d \geq (1-\mu) (\kb{\psi_{x'}} - e^\epsilon \kb{\psi_x})
        \\& \Leftrightarrow I_d \geq \frac{d(\mu -1)}{\mu(e^\epsilon-1)}(e^\epsilon \kb{\psi_x} - \kb{\psi_{x'}}).
    \end{align}
    For given $\ket{\psi_x}$ and $\ket{\psi_{x'}}$, we can choose a suitable orthonormal basis of $\calH_d$ which gives coordinate representations $\ket{\psi_x} = (1,0,0,\ldots,0)^{\top}$ and $\ket{\psi_{x'}} = (\sqrt{c_{x,x'}},\sqrt{1-c_{x,x'}},0,\ldots,0)^\top$.
    With such a basis, the upper-left $2 \times 2$ submatrix of the matrix representation of $\frac{1}{e^\epsilon-1}(e^\epsilon \kb{\psi_x} - \kb{\psi_{x'}})$ is
    {\revv{\begin{equation}
        \tau_{x,x'} = \frac{1}{e^\epsilon-1}\begin{pmatrix}
            e^\epsilon - c_{x,x'} & \omega(c_{x,x'})
            \\ \omega(c_{x,x'}) & c_{x,x'}-1
            \end{pmatrix},
    \end{equation}}}
    and all other entries are zeros, where $\omega(c_{x,x'}) := -\sqrt{c_{x,x'}(1-c_{x,x'})}$.
    Thus, we have
    \begin{align}\label{eq:QLDP_pf_1}
        e^\epsilon T_x \geq T_{x'} &\Leftrightarrow I_2 \geq \frac{d(\mu-1)}{\mu} \tau_{x,x'}.
    \end{align}

    For any $2 \times 2$ matrix $\tau$, its eigenvalues are given by
    \begin{equation}
        \frac{\mathrm{Tr}(\tau) \pm \sqrt{\mathrm{Tr}(\tau)^2 -4 \mathrm{Det}(\tau)}}{2}.
    \end{equation}
    Because $\mathrm{Tr}(\tau_{x,x'}) = 1$ and $\mathrm{Det}(\tau_{x,x'}) = \frac{c_{x,x'}-1}{4 \sinh^2(\epsilon/2)}$, the eigenvalues of $\tau_{x,x'}$ are
    \begin{equation}
        g^{\pm}_{x,x'} := \frac{1 \pm \sqrt{1+ \frac{1-c_{x,x'}}{\sinh^2(\epsilon/2)}}}{2}.
    \end{equation}
    Combining with \eqref{eq:QLDP_pf_1}, the privacy mechanism satisfies $\epsilon$-QLDP if and only if for all $x,x' \in [v]$ and $g = g_{x,x'}^\pm$,
    \begin{equation}
        1 \geq \frac{d(\mu-1)}{\mu}g.
    \end{equation}
    Because $g^-_{x,x'} \leq 0$ and $g^+_{x,x'}\geq 1$, the above condition is equivalent to
    \begin{equation}
        \forall x,x',\quad \mu \in \left[ \frac{dg^-_{x,x'}}{dg^-_{x,x'}-1}, \frac{dg^+_{x,x'}}{dg^+_{x,x'}-1} \right].
    \end{equation}
    The {\revv remaining} part concerning $g_*^{\pm}$ follows from the fact that $dg/(dg-1)$ is decreasing in $g$ for $g \leq 0$ and $g \geq 1$. \hfill $\blacksquare$

\subsection{Proof of Proposition~\ref{prop:opt_achiev_utility}}\label{sec:Q_achiev_pf}

    For $\eta = 1$, we have $\rho_{h,1} = T_{h}$.
    Then, we have
    \begin{align}\label{eq:M1_pf_1_start}
        \max\limits_{\mu} & {\revv{S^1(T) = \max\limits_{\mu} \min\limits_{h\neq h'}C(T_{h},T_{h'})}}
        \\& = -\log \left( \min\limits_{\mu} \max\limits_{h\neq h'} \min\limits_{s \in [0,1]} \mathrm{Tr}(T_h^s T_{h'}^{1-s})\right).
    \end{align}
    Considering the spectral decomposition of $T_h$, we have that
    \begin{equation}
        T_h^s = \left(\frac{\mu}{d} \right)^s I_d + \left( \left(\frac{\mu}{d}+1-\mu\right)^s - \left(\frac{\mu}{d} \right)^s \right) \kb{\psi_h}.
    \end{equation}
    By applying {\rev the AM-GM inequality} and its equality condition, we have
    \begin{equation}
        \min\limits_{s \in [0,1]}\mathrm{Tr}\left(T_h^s T_{h'}^{1-s}\right) = G(c_{h,h'},d,\mu), \label{eq:M1_pf_1}
    \end{equation}
    where $G$ is defined in \eqref{eq:large_G}.
    By calculating partial derivatives, it can be shown that
    \begin{equation}\label{eq:M1_pf_concave}
        \frac{(d-2)\mu + 2\sqrt{\mu(d-(d-1)\mu)}}{d}
    \end{equation}
    is concave in $\mu$ and its maximum value $1$ is attained at $\mu = 1$.
    Thus, the right-hand side of \eqref{eq:M1_pf_1} is increasing in $c_{h,h'} \in [0,1]$.
    Accordingly, we have
    \begin{equation}\label{eq:M_1_pf_end}
        \max\limits_{h\neq h'} \min\limits_{s \in [0,1]} \mathrm{Tr}(T_h^s T_{h'}^{1-s}) = G(c^*,d,\mu),
    \end{equation}
    where $c^* = \max\limits_{h \neq h'} c_{h,h'}$.

    Next, again by the concavity of \eqref{eq:M1_pf_concave}, the minimizer $\mu$ that minimizes the above is one of the endpoints of \eqref{eq:QLDP_mu}, $\mu_*$ or $\mu_*^+ = \frac{dg^+_*}{dg^+_*-1}$.
    We prove that actually $\mu=\mu_*$ is a minimizer, by showing
    \begin{multline}
        \frac{(d-2)\mu_* + 2\sqrt{\mu_*(d-(d-1)\mu_*)}}{d} \\ \leq \frac{(d-2)\mu_*^+ + 2\sqrt{\mu_*^+(d-(d-1)\mu_*^+)}}{d}. \label{ineq:compMinMaxMu}
    \end{multline}
    Substituting $\mu = \frac{dg}{dg-1}$ in \eqref{eq:M1_pf_concave} gives
    \begin{multline}
        \frac{(d-2)\mu + 2\sqrt{\mu(d-(d-1)\mu)}}{d}|_{\mu = \frac{dg}{dg-1}} \\ = \frac{(d-2)g}{dg-1} + \frac{2}{|dg-1|}\sqrt{g(g-1)}.
    \end{multline}
    Also, we can write
    \begin{align}
        g_*^- = -\alpha, \quad g_*^+ = 1+\alpha, 
    \end{align}
    where
    \begin{align}
        \alpha = \frac{\sqrt{1+ \frac{1-{\revv{c_{*}}}}{\sinh^2(\epsilon/2)}}-1}{2} \geq 0.
    \end{align}
    Hence, \eqref{ineq:compMinMaxMu} is equivalent to
    \begin{multline}
        \frac{(d-2)\alpha}{d\alpha+1} + \frac{2}{d\alpha+1}\sqrt{\alpha(\alpha+1)} \\ \leq \frac{(d-2)(1+\alpha)}{d\alpha+d-1}+\frac{2}{d\alpha+d-1}\sqrt{\alpha(\alpha+1)}.
    \end{multline}
    Multiplying both sides by $(d\alpha+1)(d\alpha+d-1) > 0$, {\rev the above is equivalent to}
    \begin{align}
        2(d-2)\sqrt{\alpha(\alpha+1)} \leq (d-2)(2\alpha+1). \label{ineq:compMinMaxMuFinal}
    \end{align}
    {\rev Since $d \geq 2$ and $(2\alpha+1)^2-(2\sqrt{\alpha(\alpha+1)})^2=1>0$, we have \eqref{ineq:compMinMaxMuFinal}, which implies \eqref{ineq:compMinMaxMu} and \eqref{eq:Q_PUT_eq}.}
    
    For the last part of the proof, {\rev the Taylor approximation
    of $\mu_*$ around $\epsilon=0$ gives
    \begin{equation}\label{eq:mu_star_approx}
        \mu_* =  1- \frac{\epsilon}{d\sqrt{1-c_*}} + O(\epsilon^2).
    \end{equation}
    Substituting the above into the Taylor approximation of $G(c^*,d,\mu_*)$ around $\mu_*=1$ yields
    \begin{align}
        \!\!\! G(c^*,d,\mu_*) &= 1 - \frac{d(1-c^*)}{4}(1-\mu_*)^2 + O((1-\mu_*)^3)
        \\& = 1 - \frac{1-c^*}{4d(1-c_*)}\epsilon^2 + O(\epsilon^3).
    \end{align}
    Since $-\log (1-x) = x + O(x^2)$ and $c_* \leq c^*$, we get \eqref{eq:Q_PUT_approx}.
    }
    It is trivial that the inequality in \eqref{eq:Q_PUT_approx} becomes equality if and only if $c_{x,x'}$ is constant. \hfill $\blacksquare$

\subsection{Proof of Proposition~\ref{prop:utility_cl_form}}\label{sec:Q_achiev_proposed_pf}
    Since we use SIC states, $c_{h,h'} = |\inp{\psi_h}{\psi_{h'}}|^2 = 1/(d+1)$ for all $h \neq h'$.
    The closed-form expression of $S^1(T^*)$ can be calculated by following \eqref{eq:M1_pf_1_start}--\eqref{eq:M_1_pf_end} with $\mu = \mu_*$ and $c_{h,h'} = 1/(d+1)$.
    
    Now, let $v= d^2$ for some $d \in \mathbb{N}_{\geq 2}$.
    Then, the calculation of $S^\eta(T^*)$ is similar to the case $S^1(T^*)$ because the overall overlap between all states in SIC states behaves like applying an additional depolarizing noise.
    Precisely, we have
    \begin{align}
        \frac{1}{d^2}\sum\limits_{x \in [d^2]} T^*(x) &= \frac{\mu_*}{d} I_d + \frac{1-\mu_*}{d^2} \sum\limits_{x \in [d^2]} \kb{\psi_x}
        \\& = \frac{\mu_*}{d} I_d + \frac{1-\mu_*}{d} I_d = \frac{1}{d}I_d, \label{eq:use_res_id}
    \end{align}
    where \eqref{eq:use_res_id} follows from \eqref{eq:SIC_POVM_res_id}.
    Therefore,
    \begin{align}
        & \rho_{h,\eta} = \sum\limits_{x \in [d^2]} P^{h,\eta}_x {\rev{T^*(x)}}
        \\& = \eta T^*(h) + \frac{1-\eta}{d^2} \sum\limits_{x \in [d^2]} {\rev{T^*(x)}}
        \\& = \frac{\eta\mu_*}{d} I_d + \eta(1-\mu_*) \kb{\psi_h} + \frac{1-\eta}{d}I_d
        \\& = \frac{\mu_\eta}{d} I_d + (1-\mu_\eta) \kb{\psi_h}, 
    \end{align}
    where {\rev the second-to-last and last equalities follow from \eqref{eq:use_res_id} and $\mu_\eta = 1-\eta + \eta \mu_*$, respectively.}
    Hence, we obtain \eqref{eq:S_eta} by replacing $\mu_*$ with $\mu_\eta$ in \eqref{eq:S_1}.

    The utility $A^\eta(T^*)$ in asymmetric testing can be calculated as
    \begin{align}
        & A^\eta(T^*)  = \min\limits_{h \in [d^2]} D\left( \rho_{h,\eta} \middle\| \frac{1}{d^2}\sum\limits_{x \in [d^2]} T^*(x) \right)
        \\& = \min\limits_{h \in [d^2]} D\left( \frac{\mu_\eta}{d} I_d + (1-\mu_\eta) \kb{\psi_h} \middle\| \frac{1}{d}I_d \right) \label{eq:rel_ent_cal}
    \end{align}
    where the second equality follows from \eqref{eq:use_res_id}.
    By considering matrix representations of the states in \eqref{eq:rel_ent_cal} with respect to an orthonormal basis that contains $\ket{\psi_h}$, we can calculate the relative entropy in a classical manner, i.e.,
    \begin{align}
        A^\eta(T^*)  = D\left( \frac{\mu_\eta}{d} \mathbbm{1}_d + (1-\mu_\eta) \mathbf{e}_h \middle\| \frac{1}{d}\mathbbm{1}_d \right),
    \end{align}
    where $\mathbf{e}_h$ denotes the $h$-th standard basis vector of length $d$.
    A straightforward calculation of the relative entropy yields the desired result. \hfill $\blacksquare$

\subsection{Proof of the converse part of Proposition~\ref{prop:CPUT}}\label{sec:CPUT_conv_pf}
    
    \subsubsection{Symmetric testing}
    
    Since Chernoff information satisfies the DPI, it is enough to maximize the utility $S^\eta(q)$ over extremal $\epsilon$-LDP mechanisms $q$ defined in Definition~\ref{def:extremal_mech}.
    The utility of an extremal $\epsilon$-LDP mechanism $q$ can be bounded as
    \begin{align}
        &S^\eta (q) = -\log \left( \max\limits_{h \neq h'} \min\limits_{s \in [0,1]} \sum\limits_y (\tilde{q}_{h,\eta})_y^s (\tilde{q}_{h',\eta})_y^{1-s} \right)
        \\& {\rev{\leq}} -\log \left( \max\limits_{h\neq h'} \min\limits_{s \in [0,1]} \sum\limits_{x,x'} P^{h,\eta}_x P^{h',\eta}_{x'} \sum\limits_y q_{xy}^s q_{x'y}^{1-s} \right)
        \\& \leq -\log \left( \max\limits_{h\neq h'} \sum\limits_{x,x'} P^{h,\eta}_x P^{h',\eta}_{x'} \min\limits_{s \in [0,1]}  \sum\limits_y q_{xy}^s q_{x'y}^{1-s} \right).
    \end{align}
    Here, we solve the minimization over $s$.
    Note that
    \begin{align}
        {\revv{{\revv{\chi_{h,h'}}}}}(s)&:= \sum\limits_y {\rev{q_{hy}^s q_{h'y}^{1-s}}}
        \\& = \sum\limits_y \theta_y \left( S^{(v)}_{hy} \right)^s \left(S^{(v)}_{h'y}\right)^{1-s}
    \end{align}
    is convex in $s$.
    Thus, if ${\pdv{{\revv{\chi_{h,h'}}}}{s}}(1/2) = 0$, then $\min_{s \in [0,1]}{\revv{\chi_{h,h'}}}(s) = {\revv{\chi_{h,h'}}}(1/2)$.
    In fact, this is true because
    \begin{align}
        {\pdv{{\revv{\chi_{h,h'}}}}{s}}(1/2) &= \sum\limits_y \theta_y \sqrt{\left( S^{(v)}_{hy} \right) \left(S^{(v)}_{h'y}\right)} \log \frac{S^{(v)}_{hy}}{S^{(v)}_{h'y}}
        \\& = \epsilon e^{\epsilon/2}  \left( \sum\limits_{y\in A^{e^\epsilon,1}_{h,h'}} \theta_y - \sum\limits_{y\in A^{1,e^\epsilon}_{h,h'}} \theta_y \right),
    \end{align}
    where
    \begin{equation}\label{eq:set_A}
        A_{h,h'}^{c,c'} := \left\{y : S^{(v)}_{hy} = c, S^{(v)}_{h'y} = c'\right\}.
    \end{equation}
    The following lemma proves ${\pdv{{\revv{\chi_{h,h'}}}}{s}}(1/2) = 0$, whose proof is at the {\rev end} of this subsection.
    \begin{lemma}\label{lem:theta_sum}
        Suppose $S^{(v)} \theta = \mathbbm{1}_{v}$.
        Then, for all $h \neq h'$ and $\epsilon > 0$, we have
        \begin{equation}
            \sum\limits_{y\in A^{e^\epsilon,1}_{h,h'}} \theta_y = \sum\limits_{y\in A^{1,e^\epsilon}_{h,h'}} \theta_y.
        \end{equation}
    \end{lemma} 
    
    Since $\min_{s \in [0,1]}{\revv{\chi_{h,h'}}}(s) = {\revv{\chi_{h,h'}}}(1/2)$, we have
    \begin{align}
        & \max\limits_{h\neq h'} \sum\limits_{x,x'} P^{h,\eta}_x P^{h',\eta}_{x'} \min\limits_{s \in [0,1]}  \sum\limits_y q_{xy}^s q_{x'y}^{1-s} \nonumber
        \\& = \max\limits_{h\neq h'}  \sum\limits_{x,x'} P^{h,\eta}_x P^{h',\eta}_{x'} \sum\limits_y  \theta_y \sqrt{  S_{xy}^{(v)}  S_{x'y}^{(v)} } 
        \\& \geq \frac{1}{v(v-1)} \sum\limits_{h\neq h'}  \sum\limits_{x,x'} P^{h,\eta}_x P^{h',\eta}_{x'} \sum\limits_y  \theta_y \sqrt{  S_{xy}^{(v)}  S_{x'y}^{(v)} }.
    \end{align}
    By dividing the summation over $x,x'$ into two cases, $x\neq x'$ and $x=x'$, we have  
    \begin{align}
        \sum\limits_{h\neq h'} &  \sum\limits_{x,x'}  P^{h,\eta}_x P^{h',\eta}_{x'} \sum\limits_y  \theta_y \sqrt{  S_{xy}^{(v)}  S_{x'y}^{(v)} }.
        \\ & = \sum\limits_{h\neq h'}  \sum\limits_{x} P^{h,\eta}_x P^{h',\eta}_{x} \sum\limits_y  \theta_y S_{xy}^{(v)}
        \\& \quad\quad \nonumber + \sum\limits_y  \theta_y  \sum\limits_{x\neq x'} \sum\limits_{h\neq h'} P^{h,\eta}_x P^{h',\eta}_{x'} \sqrt{  S_{xy}^{(v)}  S_{x'y}^{(v)} }
        \\& =  \sum\limits_{h\neq h'} \langle P^{h,\eta},P^{h',\eta} \rangle \label{eq:inp_hyp}
        \\& \quad\quad \nonumber + \sum\limits_y  \theta_y  \sum\limits_{x\neq x'} \sum\limits_{h\neq h'} P^{h,\eta}_x P^{h',\eta}_{x'} \sqrt{  S_{xy}^{(v)}  S_{x'y}^{(v)} },
        \\& = (1-\eta^2)(v-1)
        \\& \nonumber \quad\quad + \frac{v+ \eta^2 -1}{v} \sum\limits_y  \theta_y  \sum\limits_{x\neq x'}  \sqrt{  S_{xy}^{(v)}  S_{x'y}^{(v)} }.
    \end{align}
    where \eqref{eq:inp_hyp} follows from $\sum_y \theta_y S_{xy}^{(v)} = \sum_y q_{xy} = 1$, and the last equality follows from simple calculations based on the definition of $P^{h,\eta}$.
    Now, we minimize the last term over extremal $\epsilon$-LDP mechanisms (i.e., minimize over $\theta$) based on {\revv techniques similar to those used in the proof of} \cite[Lem.~5.3]{yoshidaMathematicalComparisonClassical2025}.
    Define
    \begin{align}
        \Omega(y) &:= \left\{x : S^{(v)}_{xy} = e^\epsilon \right\},
        \\ \beta_k &:= \sum\limits_{y:|\Omega(y)|=k} \theta_y,
        \\ w_k &:= f(v,k,\epsilon) \beta_k.
    \end{align}
    Note that the constraint $S^{(v)} \theta = \mathbbm{1}_{v}$ implies $\sum\limits_{k=0}^v w_k = 1$.
    Accordingly, {\rev{$w = (w_0,\ldots,w_v)$}} is a probability vector.
    Then,
    \begin{align}
        &\sum\limits_y  \theta_y  \sum\limits_{x\neq x'} \sqrt{  S_{xy}^{(v)}  S_{x'y}^{(v)} } \nonumber
        \\&= \sum\limits_{k=0}^v \sum\limits_{y: |\Omega(y)|=k} \theta_y \sum\limits_{x \neq x'} \sqrt{S^{(v)}_{xy} S^{(v)}_{x'y}}
        \\&= \sum\limits_{k=0}^v w_k \frac{(ke^{\epsilon/2} + v-k)^2 - (ke^\epsilon +v-k)}{f(v,k,\epsilon)}. \label{eq:C_PUT_S_prob_vec_min}
    \end{align}
    Hence, the minimum of \eqref{eq:C_PUT_S_prob_vec_min} over {\revv{$w$}} is achieved when a probability vector $w$ is a point mass.
    Therefore,
    \begin{multline}
        \min\limits_{\theta \in \mathbb{R}_{\geq 0}^{2^v}: S^{(v)}\theta = \mathbbm{1}_v} \sum\limits_y  \theta_y  \sum\limits_{x\neq x'} \sqrt{  S_{xy}^{(v)}  S_{x'y}^{(v)} } 
        \\ =  \min\limits_{k\in[0:v]}\frac{(ke^{\epsilon/2} + v-k)^2 - (ke^\epsilon +v-k)}{f(v,k,\epsilon)}
    \end{multline}
    By combining all the preceding calculations and performing straightforward algebraic manipulations, we obtain the desired result given in \eqref{eq:CPUT_S_UB}.

    \subsubsection{Asymmetric testing}
    
    Since relative entropy satisfies the DPI, it is enough to maximize the utility {\rev{$A^\eta(q)$}} over extremal $\epsilon$-LDP mechanisms $q$ defined in Definition~\ref{def:extremal_mech}.
    {\rev Similarly} to the proof of \eqref{eq:CPUT_S_UB}, we divide the summation over $y$ and bound the utility as
    \begin{align}
        A^\eta (q) &= 
        \min\limits_h \sum\limits_y (\tilde{q}_{h,\eta})_y \log \frac{(\tilde{q}_{h,\eta})_y}{(\tilde{q}_{0})_y}
        \\& = \min\limits_h \sum\limits_{k=0}^v \sum\limits_{y: |\Omega(y)|=k} (\tilde{q}_{h,\eta})_y \log \frac{(\tilde{q}_{h,\eta})_y}{(\tilde{q}_{0})_y}
        \\& \leq \frac{1}{v}\sum\limits_{k=0}^v \sum\limits_{y: |\Omega(y)|=k}\sum\limits_{h \in [v]} (\tilde{q}_{h,\eta})_y \log \frac{(\tilde{q}_{h,\eta})_y}{(\tilde{q}_{0})_y}.
    \end{align}
    If $|\Omega(y)|=k$, then we have 
    \begin{align}
        (\tilde{q}_{h,\eta})_y &=  \begin{cases} \theta_y \Delta_1 & \text{if } S^{(v)}_{hy}=e^\epsilon
        \\ \theta_y \Delta_2 & \text{if } S^{(v)}_{hy}=1 \end{cases},
        \\ (\tilde{q}_{0})_y  &= \theta_y f(v,k,\epsilon).
    \end{align}
    Then, if $|\Omega(y)|=k$, we get
    \begin{align}
        \sum\limits_{h \in [v]} (\tilde{q}_{h,\eta})_y \log \frac{(\tilde{q}_{h,\eta})_y}{(\tilde{q}_{0})_y}  = F(v,k,\epsilon)\theta_y.
    \end{align}
    Hence,
    \begin{equation}
        A^\eta(q) \leq \frac{1}{v}\sum\limits_{k=0}^v w_k \frac{F(v,k,\epsilon)}{f(v,k,\epsilon)}.
    \end{equation}
    Since the maximum of the above is achieved when a probability vector $w$ is a point mass, we get the desired result
    \begin{equation}
        A^\eta(q) \leq \frac{1}{v}\max\limits_{k \in [0:v]} \frac{F(v,k,\epsilon)}{f(v,k,\epsilon)}.
    \end{equation} \hfill $\blacksquare$

    \begin{IEEEproof}[Proof of Lemma~\ref{lem:theta_sum}]
        Let $A_{h,h'}^{c,c'}$ be the set defined in \eqref{eq:set_A}.
        Then,
        \begin{align}
            1 &= \left(S^{(v)} \theta \right)_h = \sum\limits_{y \in A_{h,h'}^{e^\epsilon,e^\epsilon}} e^{\epsilon} \theta_y + \sum\limits_{y \in A_{h,h'}^{e^\epsilon,1}} e^{\epsilon} \theta_y
            \\& \quad\quad\quad\quad\quad\quad\quad+ \sum\limits_{y \in A_{h,h'}^{1,e^\epsilon}} \theta_y
            + \sum\limits_{y \in A_{h,h'}^{1,1}} \theta_y \nonumber
            \\& = \left(S^{(v)} \theta \right)_{h'} = \sum\limits_{y \in A_{h,h'}^{e^\epsilon,e^\epsilon}} e^{\epsilon} \theta_y + \sum\limits_{y \in A_{h,h'}^{e^\epsilon,1}} \theta_y
            \\& \quad\quad\quad\quad\quad\quad\quad + \sum\limits_{y \in A_{h,h'}^{1,e^\epsilon}} e^\epsilon \theta_y \nonumber
            + \sum\limits_{y \in A_{h,h'}^{1,1}} \theta_y.
        \end{align}
        Therefore, we get the desired result because
        \begin{equation}
            (e^\epsilon-1)\left(\sum\limits_{y \in A_{h,h'}^{e^\epsilon,1}}  \theta_y - \sum\limits_{y \in A_{h,h'}^{1,e^\epsilon}} \theta_y \right) = 0.
        \end{equation}
    \end{IEEEproof}

\subsection{Proof of the achievability part of Proposition~\ref{prop:CPUT}}\label{sec:CPUT_achieve_pf}
    Let $q$ be a $(v,k,\lambda,\epsilon)$-block design mechanism.
    First, consider a symmetric testing scenario at $\eta = 1$.
    Since $(\tilde{q}_{h,1})_y = q_{hy}$, we have
    \begin{align}
        S^1(q) &= -\log \left( \max\limits_{h \neq h'} \min\limits_{s \in [0,1]} \sum\limits_y q_{hy}^s q_{h'y}^{1-s}\right).
    \end{align}
    If $h \neq h'$, the combinatorial structure of a BIBD implies
    \begin{align}
        &\sum\limits_y q_{hy}^s q_{h'y}^{1-s} \nonumber
        \\&= \frac{ \lambda e^\epsilon + (r-\lambda) \left(e^{\epsilon s} + e^{\epsilon(1-s)}\right) + b-2r+\lambda }{re^\epsilon+b-r}.
    \end{align}
    Here, the AM-GM inequality implies that this value can be minimized when $s=1/2$.
    Hence, we get
    \begin{align}
        S^1(q) &= -\log \left(
        \frac{\lambda e^\epsilon + 2(r-\lambda) e^{\epsilon/2} + b-2r+\lambda}{re^\epsilon+b-r} \right)
        \\& = -\log \left( 1-\frac{(e^{\epsilon/2}-1)^2}{v-1} \cdot \frac{k(v-k)}{ke^\epsilon+v-k}\right),
    \end{align}
    where the last equality follows from the identities in \eqref{eq:BD_necc}.
    Finally, we obtain $S^1(q) = \overline{S^1_\mathrm{C}}(v,\epsilon)$ by choosing $k \in [0:v]$ which maximizes the above value. 

    Now, consider asymmetric testing scenario and $\eta \in [0,1]$ and recall the notations in \eqref{eq:not_A_start}--\eqref{eq:not_A_end}.
    Note that by Definition~\ref{def:BD_mech},
    \begin{align}
        (\tilde{q}_{h,\eta})_y &= \begin{cases} \frac{\Delta_1}{re^\epsilon + b-r} & \text{if } h \in e_y
        \\ \frac{\Delta_2}{re^\epsilon + b-r} & \text{if } h \not\in e_y
        \end{cases},
        \\ (\tilde{q}_{0})_y & = \frac{f(v,k,\epsilon)}{re^\epsilon + b-r}.
    \end{align}
    By Definition~\ref{def:BIBD}, for any $h \in [v]$, there are exactly $r$ blocks that contain $h$.
    Hence, the utility of a block design mechanism $q$ is calculated as 
    \begin{align}
        A^\eta(q) &= \min\limits_h \sum\limits_y (\tilde{q}_{h,\eta})_y\log \frac{(\tilde{q}_{h,\eta})_y}{(\tilde{q}_{0})_y}
        \\& = \frac{r\Delta_1}{re^\epsilon + b-r} \log \frac{\Delta_1}{f(v,k,\epsilon)}
        \\& \quad\quad\quad\quad\quad\quad + \frac{(b-r)\Delta_2}{re^\epsilon + b-r} \log \frac{\Delta_2}{f(v,k,\epsilon)} \nonumber
        \\& = \frac{k\Delta_1}{ke^\epsilon + v-k} \log \frac{\Delta_1}{f(v,k,\epsilon)} \label{eq:vrbk_usage}
        \\& \quad\quad\quad\quad\quad\quad + \frac{(v-k)\Delta_2}{ke^\epsilon + v-k} \log \frac{\Delta_2}{f(v,k,\epsilon)}\nonumber
        \\& = \frac{F(v,k,\epsilon)}{vf(v,k,\epsilon)},
    \end{align}
    where \eqref{eq:vrbk_usage} follows from the identity $vr=bk$.
    Finally, we obtain the desired result by choosing $k \in [0:v]$ which maximizes the above value. \hfill $\blacksquare$

\section{Conclusion \& Discussions}\label{sec:conc}
{\revv We demonstrated a quantum advantage in private hypothesis testing with respect to the optimal PUT in both symmetric and asymmetric testing. Specifically, we established this advantage for a particular class of hypotheses involving smoothed point mass distributions, under stringent privacy constraints and small discrete data alphabet sizes from $3$ to $9$.}
To establish this result, we proposed a QLDP mechanism that prepares pure states and then subjects them to a depolarizing channel.
We chose the depolarizing parameter and pure states to maximize the utility in certain cases, and the chosen states are SIC states.
In addition, we derived an upper bound on the classical PUT in symmetric testing, which is tight in certain parameter regimes, and exactly characterized the classical PUT in asymmetric testing.

{\revv An interesting direction for future work would be to show} quantum advantage for all parameters $v,\eta,\epsilon$, or for more general hypotheses, or for other private statistical inference tasks.
Regarding the first two directions, a block design mechanism \cite{parkExactlyOptimalCommunicationEfficient2024} generalized to a quantum system can be considered.
In more detail, we can consider a CQ channel whose outputs are mixtures of pure states $\{\ket{\psi_e}\}$ where the coefficients are determined by a block design (cf. Definition~\ref{def:BD_mech}) as follows:
\begin{multline}
    T_x = \frac{\mu}{r\mu + (b-r)(1-\mu)} \sum\limits_{e \in B: x \in e} \kb{\psi_e}
    \\ + \frac{1-\mu}{r\mu + (b-r)(1-\mu)} \sum\limits_{e \in B: x \not\in e} \kb{\psi_e},
\end{multline}
where $([v],B)$ is a $(v,k,\lambda)$-design.
However, there are technical challenges in calculating the utility and verifying the QLDP constraint, as specifying the spectral decomposition of a general mixture of pure states is difficult.
If there are some choices of block designs and pure states $\{\ket{\psi_e}\}$ that circumvent these difficulties, they may show quantum advantages for more general cases.
For the last direction, widely used inference tasks such as parametric estimation and mean estimation would be considered.

{\rev Another interesting direction would be to precisely characterize the optimal quantum PUT.
One approach would be to extend the concept of extremal LDP mechanisms \cite{kairouzExtremalMechanismsLocal2016b}, which {\revv has} been utilized in characterizing the optimal classical PUT, to QLDP mechanisms.
Obtaining a simpler characterization of the optimal quantum PUT would allow us to clarify the regimes where quantum advantage exists or does not exist by comparing with the optimal classical PUT.}

\section*{Acknowledgments}
This work was supported in part by Basic Science Research Program through the National Research Foundation of Korea (NRF) funded by the Ministry of Education (No. RS-2024-00452156), in part by the NRF grant funded by the Korea government (MSIT) (No. RS-2025-00561467, RS-2024-00408613), in part by Institute of Information \& Communications Technology Planning \& Evaluation (IITP) grant funded by the Korea government (MSIT) (No. RS-2023-00229524, Quantum Certification and Its Applications to Quantum SW (QC\&QSW); RS-2023-00215700, Trustworthy Metaverse: Blockchain-Enabled Convergence Research; RS-2025-02304540; RS-2025-25464876; RS-2025-25464616), and in part by IITP grant funded by the Korea Media and Communications Commission (KMCC) (No. RS-2026-25514346, Persona AI Model Technology Based on Accumulated Media Consumption Histories).

\bibliographystyle{IEEEtran}
\bibliography{QDPref}

\vspace{-20pt}

\begin{IEEEbiography}[{\includegraphics[width=1in,height=1.25in,clip,keepaspectratio]{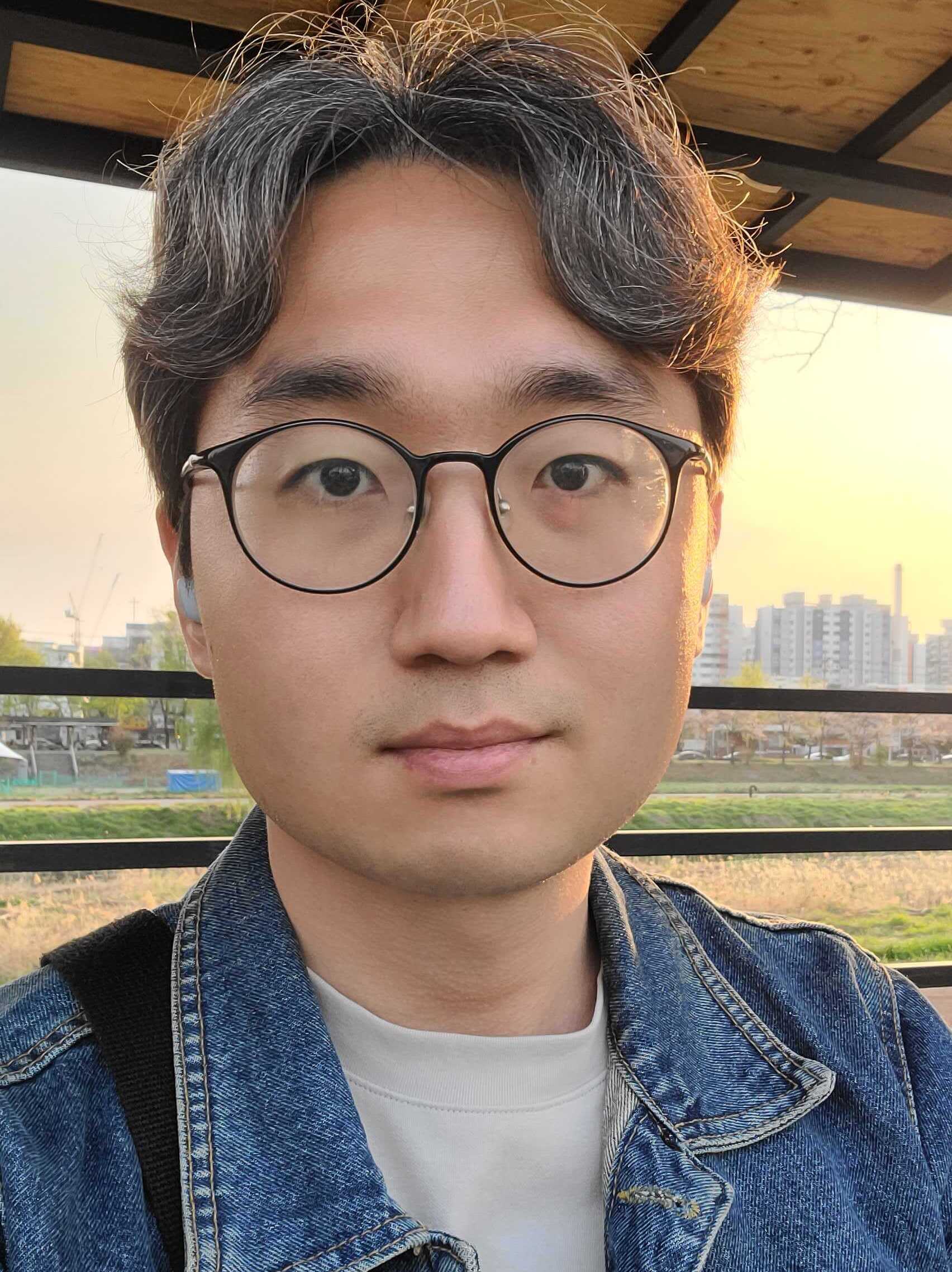}}]{Seung-Hyun Nam} (Member, IEEE) received the B.S. and M.S. degrees in Electrical Engineering from Pohang University of Science and Technology (POSTECH), Pohang, South Korea, in 2018 and 2020, respectively, and Ph.D. degree in Electrical Engineering from Korea Advanced Institute of Science and Technology (KAIST), Daejeon, South Korea, in 2024.
He is currently a post-doctoral researcher with Information \& Electronics Research Institute, KAIST.
His research interests include (both classical and quantum) information theory, statistical inference, differential privacy, and information theoretic security.
\end{IEEEbiography}

\begin{IEEEbiography}[{\includegraphics[width=1in,height=1.25in,clip,keepaspectratio]{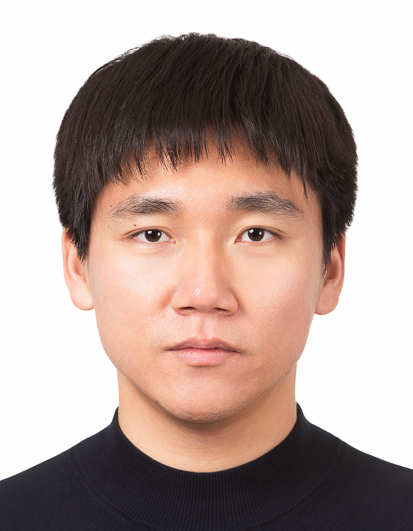}}]{Hyun-Young Park} (Member, IEEE) received the B.S. (valedictorian) and Ph.D. degrees in Electrical Engineering from Korea Advanced Institute of Science and Technology (KAIST), Daejeon, South Korea, in 2021 and 2026, respectively.
He is currently a post-doctoral researcher with the School of Electrical Engineering, KAIST. His research interests include information theory, differential privacy, and quantum information theory.
\end{IEEEbiography}

\begin{IEEEbiography}[{\includegraphics[width=1in,height=1.25in,clip,keepaspectratio]{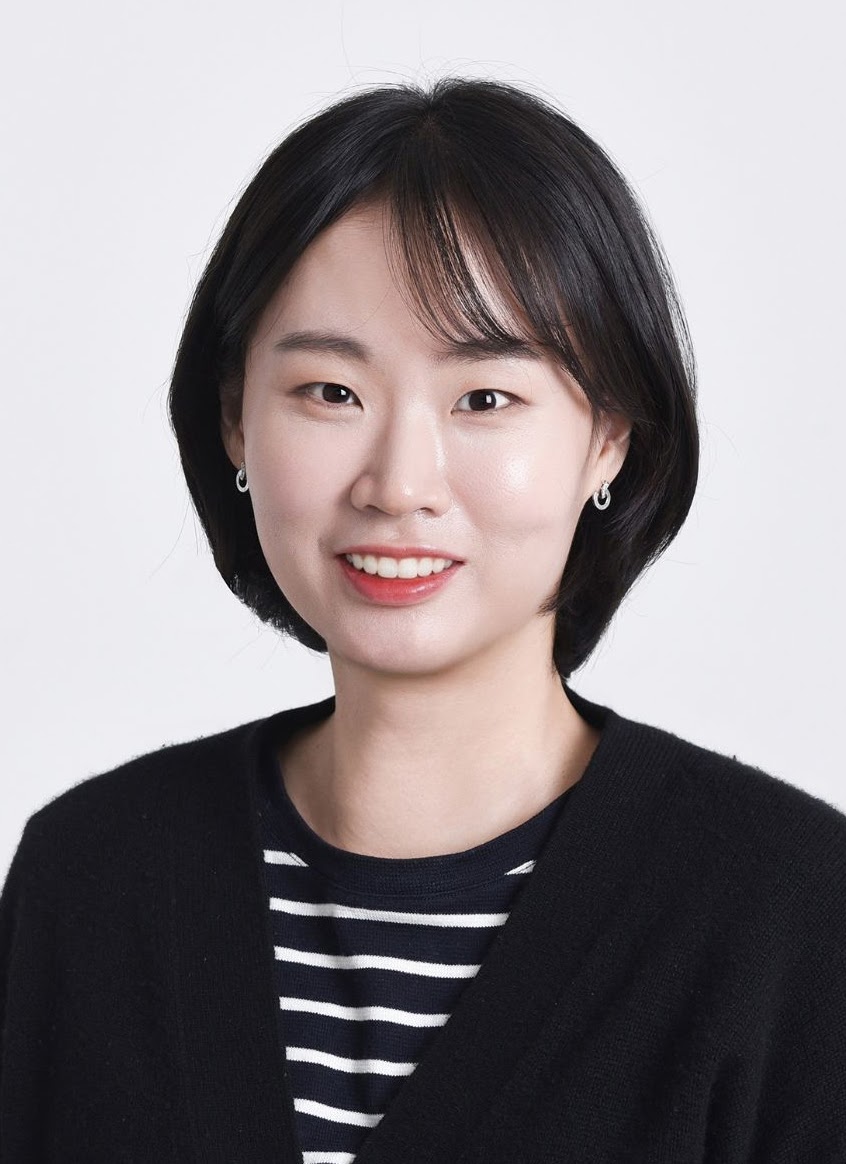}}]
{Si-Hyeon Lee} (Senior Member, IEEE) received the B.S. (summa cum laude) and Ph.D. degrees in electrical engineering from the Korea Advanced Institute of Science and Technology (KAIST), Daejeon, South Korea, in 2007 and 2013, respectively. She is currently an Associate Professor with the School of Electrical Engineering, KAIST. She was a Postdoctoral Fellow with the Department of Electrical and Computer Engineering, University of Toronto, Toronto, Canada, from 2014 to 2016, and an Assistant Professor with the Department of Electrical Engineering, Pohang University of Science and Technology (POSTECH), Pohang, South Korea, from 2017 to 2020. Her research interests include information theory, wireless communications, statistical inference, and machine learning. 
She was an IEEE Information Theory Society Distinguished Lecturer (2024-2025), a TPC Co-Chair of IEEE Information Theory Workshop 2024, and  a Guest Editor for {\scshape IEEE Journal on Selected Areas in Communications} (Special issue on Secure Communication, Sensing, and Computation in Future Intelligent Wireless Networks). She is currently an Associate Editor for {\scshape IEEE Transactions on Information Theory} and {\scshape IEEE Transactions on Communications}.
\end{IEEEbiography}

\begin{IEEEbiography}[{\includegraphics[width=1in,height=1.25in,clip,keepaspectratio]{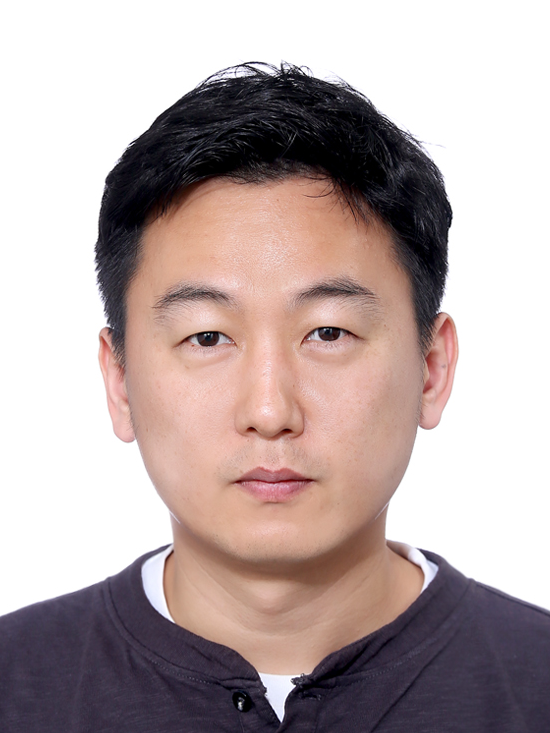}}]{Joonwoo Bae} (Member, IEEE) obtained a Ph.D. in Theoretical Physics from Universitat de Barcelona \& ICFO-Institute of Photonic Sciences, Barcelona in 2007. He has worked at the Korea Institute for Advanced Study (KIAS), Centre for
Quantum Technologies (CQT) in Singapore, the ICFO, Freiburg Institute for Advanced Studies (FRIAS) as a Junior Fellow, and Hanyang University. He is currently with the School of Electrical Engineering, Korea Advanced Institute of Science and Technology (KAIST). His research interests include secure quantum communication, entanglement applications, open quantum systems, quantum foundations, and their practical applications.
\end{IEEEbiography}

\vfill

\end{document}